\documentclass[lettersize,journal]{IEEEtran}
\usepackage{amsmath,amsfonts}
\usepackage{algorithm}
\usepackage{array}
\usepackage{textcomp}
\usepackage{stfloats}
\usepackage{url}
\usepackage{verbatim}
\usepackage{graphicx}
\usepackage{cite}
\usepackage{amsmath}
\usepackage{amssymb}
\usepackage{xspace}
\usepackage[algo2e]{algorithm2e} 
\usepackage{setspace}
\usepackage{latexsym,fancyhdr,url}
\usepackage{enumerate}
\usepackage{graphics}
\usepackage{xparse} 
\usepackage{xspace}
\usepackage{multirow}
\usepackage{csvsimple}
\usepackage{tabularx}
\usepackage{adjustbox}
\usepackage{booktabs}
\usepackage{mathtools}
\usepackage{balance}
\usepackage{algpseudocode}
\usepackage{nicefrac}
\usepackage{siunitx}
\usepackage{array,framed}
\usepackage{listings}
\usepackage{enumitem}
\usepackage{subfigure}
\usepackage[normalem]{ulem}
\usepackage{xcolor}
\usepackage{multicol}

\usepackage{
  tikz,
  subcaption}

\newcommand\sysname{DEXO\xspace}

\newcommand*\circled[1]{\tikz[baseline=(char.base)]{
            \node[shape=circle,draw,inner sep=1pt] (char) {#1};}}

\newcommand\rev[1]{\textcolor{black}{#1}}

\newcommand{\revtwo}[1]{\textcolor{black}{#1}}
\newcommand\redout{\bgroup\markoverwith{\textcolor{red}{\rule[.5ex]{2pt}{0.6pt}}}\ULon}

\usepackage{pifont}
\newcommand{\cmark}{\textcolor{green}{\ding{51}}}
\newcommand{\xmark}{\textcolor{red}{\ding{55}}}

\newtheorem{definition}{Definition}
\newtheorem{theorem}{Theorem}
\newtheorem{lemma}{Lemma}

\begin{document}

\title{\sysname: A Secure and Fair Exchange Mechanism for Decentralized IoT Data Markets}

\author{
    \IEEEauthorblockN{
        Yue Li\IEEEauthorrefmark{1}, 
        Ifteher Alom\IEEEauthorrefmark{1}, 
        Wenhai Sun\IEEEauthorrefmark{2},
        Yang Xiao\IEEEauthorrefmark{1} 
    }\\
    \IEEEauthorblockA{\IEEEauthorrefmark{1}University of Kentucky, Lexington, KY, USA} (email: \{yue.li,ifteheralom,xiaoy\}@uky.edu)\\
    \IEEEauthorblockA{\IEEEauthorrefmark{2}Purdue University, West Lafayette, IN, USA} (email: whsun@purdue.edu)
    

    \thanks{
    
    This is the authors' accepted version of the published article:

    Y.~Li, I.~Alom, W.~Sun, and Y.~Xiao, ``DEXO: A Secure and Fair Exchange Mechanism for Decentralized IoT Data Markets,'' \emph{IEEE Internet of Things Journal}, vol.~12, no.~11, pp.~16095--16111, 2025. doi: 10.1109/JIOT.2025.3535671. 
  
    Copyright~(c)~2025 IEEE. Personal use of this material is permitted. However, permission
  to use this material for any other purposes must be obtained from the IEEE by sending
  a request to \texttt{pubs-permissions@ieee.org}.
  }
}



\maketitle

\begin{abstract}
Opening up data produced by the Internet of Things (IoT) and mobile devices for public utilization can maximize their economic value. Challenges remain in the trustworthiness of the data sources and the security of the trading process, particularly when there is no trust between the data providers and consumers. In this paper, we propose DEXO, a decentralized data exchange mechanism that facilitates secure and fair data exchange between data consumers and distributed IoT/mobile data providers at scale, allowing the consumer to verify the data generation process and the providers to be compensated for providing authentic data, with correctness guarantees from the exchange platform. To realize this, DEXO extends the decentralized oracle network model that has been successful in the blockchain applications domain to incorporate novel hardware-cryptographic co-design that harmonizes trusted execution environment, secret sharing, and smart contract-assisted fair exchange. For the first time, DEXO ensures end-to-end data confidentiality, source verifiability, and fairness of the exchange process with strong resilience against participant collusion. We implemented a prototype of the DEXO system to demonstrate feasibility. The evaluation shows a moderate deployment cost and significantly improved blockchain operation efficiency compared to a popular data exchange mechanism. 
\end{abstract}

\section{Introduction}
\label{sec:intro}



\IEEEPARstart{D}{ata} produced by mobile and Internet of Things (IoT) devices are widely seen as valuable assets for the knowledge-based economy, with important applications in mobile network planning, city traffic management, healthcare analytics, and more recently training foundational AI models.
Unlike the free data on the Internet, IoT data are usually proprietary and have limited trust boundaries. Sharing these data with consumers from other domains would have profound security and privacy implications. A trusted third party (TTP) such as a data broker is often required to facilitate the data acquisition from distributed providers and the sale of data to interested parties. This model, however, represents a central point of failure and is prone to privacy violations. In an infamous case, T-Mobile was found selling mobile subscribers' location data to third-party data brokers who subsequently sold them to other unauthorized parties, all without subscribers’ consent \cite{fcc-tmobile}. Similar incidents have occurred in other big telecoms \cite{fcc-att,fcc-verizon} and various data brokers \cite{Sherman-data-brokers}.

Ideally, building a data marketplace requires a secure exchange mechanism that facilitates data sales between data providers and data consumers \cite{driessen2022data}. When it comes to a marketplace for mobile/IoT data, the exchange mechanism faces four specific security challenges.
(i) \emph{End-to-end data confidentiality}: the data of interest should only be revealed to the paid consumer in a process transparent to the data provider; it remains confidential to third parties, including the facilitators of the exchange.
(ii) \emph{Fair exchange}: the exchange process should ensure that the consumer receives the data only if a payment is made to the provider; the provider receives the payment only if the consumer gets the data. The consumer should also be able to revert/abort the exchange if the data does not meet promised specifications.
(iii) \emph{Data source quality and verifiability}: the consumer should receive quality data that conforms to a pre-agreed standard and can be verified for its integrity.
(iv) \emph{Resilience}: the exchange process should not suffer from single-point failures; the above goals can be achieved even if a fraction of participants malfunction or collude. Besides security goals, the exchange mechanism should have good \emph{scalability} in data volume, due to the sheer size of data from the distributed mobile/IoT devices.

The recent rise of distributed ledger technology, represented by blockchain, and its native \emph{smart contract} functionality have offered a viable path toward the above vision. Smart contracts allow for the automatic and traceable execution of business logic between untrustful parties with the correctness and liveness enforced by the underlying blockchain consensus. In light of this, there has been extensive research leveraging smart contracts to enable mobile/IoT data marketplaces. 
One line of research leverages smart contracts as the \emph{on-chain} element of a trusted data broker that facilitates the listing and sale of data \cite{missier2017mind,suliman2018monetization,bajoudah2019toward,databroker-global,protocol2020decentralized}. The data item of interest is usually curated in the broker's \emph{off-chain} server confidentially which can also perform quality control \cite{ramachandran2018towards,sharma2020towards}; the contract encodes access control rules that determine the release of the data to the consumer upon receiving a valid bid with payment from the latter. This paradigm bears some similarities to the non-fungible token (NFT) marketplaces \cite{Rarible-Official-Website,OpenSea-Official-Website}, where the access control centers around the transfer of data ownership (by utilizing digital signatures) instead of the transfer of data item itself. In both cases, significant trust is placed in the broker's off-chain server for storing and transferring data upon contract rules, posing a risk of single-point failure. 


When it comes to data source quality control and verifiability, a popular data provision paradigm known as \emph{blockchain oracle} provides a potential solution. Blockchain oracles are third-party services designed to help a smart contract procure real-world data critical to its application logic \cite{chainlinkoracleproblem}. 
Compared to data marketplaces, blockchain oracles focus on the data provision process by building a secure channel between the data sources and a consumer contract \cite{zhang2016town}. The data items are usually collected and curated by a dedicated group of server nodes called the Decentralized Oracle Network (DON) with each node selecting its own data sources \cite{breidenbach2021chainlink,bandprotocol}. DON nodes aggregate their data on an oracle contract that serves as the query interface to potential consumer contracts.
Despite their popularity, existing DONs heavily consolidate the upstream data provision process and can only deliver data on the blockchain. This leads to data diversity and scalability (on-chain cost) challenges that hinder their applicability to data markets \cite{benligiray2020decentralized,xiao2023decentruth}.
The on-chain data delivery is also restricted to non-confidential data. 
Nonetheless, DONs provide valuable lessons and established infrastructure for securing the data supply side, which is potentially useful for building a marketplace of verified data.


In this paper, we introduce \textbf{\sysname} (\textbf{D}ecentralized data \textbf{EX}change \textbf{O}racle), a new data exchange platform designed to enable a secure and scalable marketplace for mobile/IoT data. 
\sysname extends the DON model into a decentralized data exchange platform that for the first time accomplishes the security goals of end-to-end data confidentiality, source verifiability, and fair exchange of data with strong resilience to single-point failures. 
The main infrastructure of the \sysname platform consists of a DON-like node consortium called the \textbf{\sysname Network} as the off-chain component and \textbf{\sysname Contracts} as the on-chain component. On a high level, the \sysname Network is responsible for the collection and curation of ciphertext data from data providers. A provider-specific \sysname Contract is responsible for data listing and enforcing data access control, fair exchange, and compensation.

On the data supply side, we require owners of IoT/mobile devices that can produce common sensory data to form a data-provider decentralized application (DApp), dubbed P-DApp. 
The P-DApp represents the data owners in the \sysname data market with a frontend server responsible for collecting and transporting data from each device to the \sysname Network and a dedicated \sysname Contract serving as its on-chain backend, fulfilling data listing and later data owner compensation upon a successful sale. 
To address the data quality and verifiability challenges, \sysname leverages the emerging availability of trusted execution environments (TEE) in commercial IoT/mobile devices \cite{ARMTrustZone,apple2019T2} which provides attested execution of sensitive applications. In the data generation stage, \sysname requires each data owner device to instantiate a \sysname-ratified TEE application $\mathcal{F}_{TA}$ that pre-processes raw data and sanitizes them into the required format. $\mathcal{F}_{TA}$ and its output are verifiable for \emph{execution integrity} and \emph{device authenticity} with the help of TEE's attested execution capability \cite{OP-TEE-Attestation1}. This ensures that the data originates from $\mathcal{F}_{TA}$-equipped devices instead of being mass-generated by unknown sources.

\sysname further achieves end-to-end confidentiality and resilience in the data exchange by integrating secret sharing and a fair exchange mechanism into the data generation and exchange workflow.
Besides pre-processing raw data and attaching integrity proofs, $\mathcal{F}_{TA}$  splits the data into $N$ secret shares with each share forwarded through the P-DApp to a specific \sysname node (assuming there are $N$ \sysname nodes). This ensures that each data share, even if being intercepted during transit, remains unintelligible to unauthorized entities and the plaintext data remains confidential to individual \sysname nodes. Only entities possessing a threshold fraction of secret shares ($t$ out of $N$) can reconstruct the original data. 

When a data consumer sends a purchase request to \sysname Contract for a certain data item,
each \sysname node will be engaged in a fair exchange process to deliver the corresponding data shares to the consumer. The P-DApp users should receive compensation only if the consumer obtains the correct data at the end of the exchange. This process involves an atomic execution of off-chain delivery of encrypted data shares and on-chain release of the decryption key by utilizing cryptographic commitments \cite{eckey2020optiswap}. 
Under the assumption that $F$ out of the $N$ \sysname nodes are compromised with $F<\frac{1}{2}N$ and $F<t\leq N-F$,
we prove that the original data remains confidential to individual \sysname nodes at all times and the consumer is guaranteed to reconstruct the data. \sysname also guarantees resilience against potential collision cases in that by colluding with fewer than $t$ \sysname nodes, a consumer cannot scam a P-DApp for any portion of the original data without full payment. Likewise, a P-DApp cannot scam a consumer for payment without providing the full requested data or by colluding with individual nodes who attempt to tamper with the data shares.

To sum up, we make the following contributions:
\begin{itemize}
\item We propose \sysname, a new decentralized data exchange mechanism to enable a secure and scalable marketplace for IoT and mobile data. \rev{\sysname extends the DON model and} for the first time accomplishes end-to-end data confidentiality, source verifiability, \rev{fault tolerance}, and \rev{fair exchange of off-chain data}.

\item 
\rev{\sysname's data supply side leverages TEE-based data secret sharing to realize confidential and verifiable data procurement from IoT devices. This design is of independent interest to DON services for sourcing sensitive data with origin verifiability. 
\item \sysname's data exchange side leverages smart contract-based fair exchange for delivering off-chain data from DON nodes to consumers while enforcing payments to original data sources.} Our design ensures strong guarantees of fault tolerance and collusion resistance against malicious \rev{system participants and also handles disputes between data owners and consumers.}


\item We provide a proof-of-concept implementation of \sysname, utilizing Ethereum as the smart contract platform and ARM TrustZone as the IoT device TEE platform. The experiment results illustrate that \sysname significantly outperforms existing DON solutions in on-chain gas cost per unit of data consumed, while incurring moderate off-chain execution overhead for individual data providers, demonstrating \sysname efficiency and practicality.
\end{itemize}

The remainder of the paper is organized as follows. Section \ref{sec:related-work} discusses the existing work relevant to our scheme. Section \ref{sec:system-model} describes the system models and goals. Section \ref{sec:preliminaries} introduces the building blocks necessary for constructing our scheme. Section \ref{sec:detailed-design} elaborates on the detailed design. Section \ref{sec:analyses} provides security and complexity analyses of our scheme. The implementation and evaluation results are presented in Sections \ref{sec:implementation}-\ref{sec:evaluation}, followed by the conclusion in Section \ref{sec:conclusion}.

\section{Background and Related Work}
\label{sec:related-work}



\begin{table*}
    \centering
    \small
    \caption{Comparison of Major Data Market Mechanisms and \sysname. Five criteria: (1) Data Source Verifiability---the reliability of a data source can be technically verified; (2) End-to-end Data Confidentiality---data is kept private and not exposed on public platforms like blockchains or non-consumers; (3) Decentralization---preventing system's single-point failure with strong resilience; (4) Fair Exchange of Off-chain Data---allows consumers to obtain refunds if the data does not meet promised specifications; (5) Mobile/IoT Data---support for the exchange of mobile/IoT data.}
    \begin{tabular}{lcccccc}
        \toprule
         &  & Data Source & End-to-end Data & Decentralization & Fair Exchange & Mobile/IoT \\
               Scheme  & Category & Verifiability & Confidentiality & (Fault Tolerance) & of Off-chain Data &  Data \\
        \midrule
        Ocean Protocol \cite{protocol2020decentralized} & data exchange & \xmark & \cmark & \cmark & \xmark & \cmark \\ 
        DataBroker DAO \cite{databroker-global} & data exchange & \xmark & \cmark & \cmark & \cmark & \cmark \\
        OpenSea \cite{OpenSea-Official-Website} & NFT market & provenance & \xmark & \cmark & ownership only  & \xmark \\
        Rarible \cite{Rarible-Official-Website} & NFT market & provenance & \xmark & \cmark & ownership only  & \xmark \\
        FairSwap \cite{dziembowski2018fairswap} & data exchange & \xmark & \cmark & \xmark & \cmark & \cmark \\
        OptiSwap \cite{eckey2020optiswap} & data exchange & \xmark & \cmark & \xmark & \cmark & \cmark \\
        PrivacyGuard \cite{xiao2020privacyguard} & data exchange & \xmark & \cmark & \xmark & \cmark & \cmark \\
        Town Crier \cite{zhang2016town} & oracle service & \cmark & \xmark & \xmark & \xmark & \xmark \\
        DECO \cite{zhang2020deco} & oracle service & \cmark & \cmark & \cmark & \xmark & \cmark \\
         Chainlink \cite{breidenbach2021chainlink} & oracle service & \cmark & \xmark & \cmark & \xmark & \xmark \\ \midrule
         \textbf{\sysname}& data exchange & \cmark & \cmark & \cmark & \cmark & \cmark \\  
         \bottomrule
    \end{tabular}
    \label{tab:comparison-solutions}
\end{table*}

\subsection{Smart Contracts and DApps}

Smart contracts facilitate the automated and transparent execution of business logic among parties that do not trust each other and are typically instantiated on a distributed ledger or blockchain system \cite{wood2014ethereum,eos2018whitepaper}.
DApps are web applications that leverage blockchain smart contracts for creating decentralized, self-governing, and minimum-trust business logic. A DApp usually comprises three parts: a user interface (such as a browser), a user-facing server as the \emph{frontend}, and a blockchain smart contract as the \emph{backend} \cite{chainlink2022dapp}. The user interface and frontend server function similarly to traditional web applications, while the smart contract backend is responsible for processing and storing the DApp's main transactional logic whose security is provided by the underlying blockchain consensus. This contrasts with traditional apps whose backend logic is usually handled by a centralized cloud server.
\sysname leverages the DApp format to standardize the supply side of the data market by requiring IoT device owners to join data-providing DApps that follow standardized data provision and compensation workflows.

\subsection{Decentralized Data Exchanges~}


The salient properties of smart contracts also give rise to decentralized data marketplaces that are transparent, auditable, and autonomous.
For example, smart contracts can serve as a listing platform for data that are originally stored in an off-chain data broker \cite{missier2017mind,suliman2018monetization,bajoudah2019toward}. The smart contract can encode certain access and compensation rules that determine the action of the data broker on the release of data to a paid consumer.
Ocean Protocol~\cite{protocol2020decentralized}
is a decentralized data exchange platform to facilitate data sharing and monetization to unlock data for AI. 
It relies on its tokenized service layer (with the OCEAN data token) to mediate data exchange.
DataBroker DAO~\cite{databroker-global} targets IoT sensory data, providing a decentralized marketplace for users to buy and sell data generated by IoT devices. It employs smart contracts to facilitate IoT data transactions while keeping records as the contract states.
Streamr~\cite{Streamr-White-Paper} is a decentralized, peer-to-peer platform for real-time data sharing, concentrating on creating and operating data streams. 
In comparison to \sysname, these solutions generally do not fulfill data source verification or fair exchange (except DataBroker DAO) as they place trust in the autonomous market participants and data brokers \cite{driessen2022data}.

Another popular type of data exchange is NFT marketplaces. A typical NFT marketplace, such as OpenSea~\cite{OpenSea-Official-Website} and Rarible~\cite{Rarible-Official-Website}, takes temporary ownership of the token using an escrow account and the token is transferred to the highest bidder. The exchange process involves the on-chain transfer of ownership where the seller signs the transfer-out invocation with the new owner's account address. It does not involve off-chain fair exchange since the digital object referenced by the NFT is available publicly (at least in partial form). In comparison, \sysname focuses on the value of data itself rather than ownerships. That is, \sysname needs to keep the data confidential from the \sysname marketplace before the buyer provides payment.



\subsection{Smart Contract-based Fair Exchange Protocols}

Fair exchange protocols aim to address the lack of trust between the parties of a digital trade. A fair exchange protocol should realize an atomic exchange, meaning the seller is ensured that the buyer can only receive the digital asset when the payment is received, and the buyer is ensured that the seller can only receive the payment when the digital asset is obtained, which together fulfills an atomic exchange. While it is shown that fair exchange is not possible without a TTP \cite{pagnia1999impossibility}, the emergence of blockchain-based smart contracts shows a viable solution by having a smart contract fulfilling the TTP role 
\cite{alsharif2020blockchain,lopez2020multi,campanelli2017zero,dziembowski2018fairswap,eckey2020optiswap,xiao2020privacyguard}. 

Two popular \rev{smart contract-based fair exchange} schemes are FairSwap \cite{dziembowski2018fairswap} and OptiSwap \cite{eckey2020optiswap}. They leverage cryptographic commitments to facilitate the transfer of the buyer's pre-payment to the seller and the release of the decryption key of the data asset to the buyer. In FairSwap \cite{dziembowski2018fairswap}, the \rev{seller} initially provides encrypted data along with some auxiliary information to the buyer. The buyer checks the auxiliary information and, if convinced, deposits the money into the smart contract. Once the seller has received an assurance of the payment (locked at the smart contract), the secret key is released to the blockchain, and the buyer is thus able to decipher the witness. FairSwap protocol also employs a Merkle proof-based mechanism called the proof of misbehavior (PoM) concept to deal with invalid witnesses. OptiSwap \cite{eckey2020optiswap} extends FairSwap by incorporating an interactive dispute resolution protocol executed only in pessimistic cases, thus expediting honestly performed transactions.
Specifically, the buyer and seller have a pre-determined predicate function $\phi()$ on the validity of a given data item $x$ ($\phi(x) == 1$ means valid; 0 means invalid). 
PrivacyGuard \cite{xiao2020privacyguard} achieves a similar goal by using TEE for off-chain data storage and a hash lock mechanism on a smart contract for disclosing the key. While these schemes nicely achieve data confidentiality and fair exchange goals, they do not provide functions for data source quality control or verification and also face risks of single-point failures due to centrally managed data provisioning and storage.

\rev{
Independent of the above fair exchange schemes, Hash Time-Locked Contract (HTLC) provides another paradigm for exchanging on-chain assets, particularly for enabling cross-chain atomic swaps \cite{herlihy2018atomic}. HTLC-based atomic swap protocols allow two parties to exchange native cryptocurrencies without relying on trusted third parties and guarantee atomicity---either both receive the cryptocurrencies from each other or neither does. Particularly, due to HTLC's time locking mechanism, if one party does not claim the funds within a specific time, the other party can reclaim the funds and rescind the swap.
Recent developments including MAD-HTLC \cite{tsabary2021mad} and He-HTLC \cite{wadhwa2022he} have addressed HTLC's vulnerability to bribery attacks and other strategic manipulation
by cryptocurrency miners, elevating the security of HTLC-based atomic swaps from a game-theoretical perspective.
In comparison, fair exchange schemes (including FairSwap \cite{dziembowski2018fairswap}, OptiSwap \cite{eckey2020optiswap}, and our adaptation) differ from atomic swaps in application scenarios and certain security properties.
First, atomic swap is an effective method to exchanging cryptocurrencies (or other on-chain assets) across different blockchains, whereas fair exchange enables the exchange of cryptocurrency for any off-chain digital assets.
Second, atomic swap relies on time locks to facilitate graceful exit (required for atomicity), whereas fair exchange relies on cryptographic commitments and custom-defined predicate functions to facilitate a dispute process.
Third, fair exchange supports the off-chain delivery of confidential assets, providing a unique advantage for exchanging sensitive data.
}

\subsection{Blockchain Data Oracles}
Blockchain data oracles are third-party services that transport data from external (off-chain) sources into smart contracts. 
Traditionally, blockchain oracle schemes focus on providing secure channels between smart contracts and external data sources \cite{zhang2016town,zhang2020deco,guarnizo2019pdfs,benligiraydecentralized}. Town Crier \cite{zhang2016town} extends Transport Layer Security (TLS) for establishing authenticated communication between HTTPS-enabled websites to client contracts by leveraging TEEs as a trusted intermediary. DECO \cite{zhang2020deco} achieves similar functions through multiparty computation and further provides data confidentiality by using zero-knowledge proofs (ZKP) to validate oracle events without exposing the data in plaintext. 

Popular data oracle services such as Chainlink \cite{breidenbach2021chainlink} and Band Protocol \cite{bandprotocol} have offered practical solutions to the data provisioning problem by adopting the decentralized oracle network (DON) model. The DON model stipulates that for every data query (\emph{e.g.,} latest price of a certain asset), a network of independent oracle nodes collect responses from their own selection of sources. 
To avoid a single point of failure and to reign in the varying quality of data sources, each oracle node aggregates the local responses and submits the result to an on-chain oracle contract. The contract automatically aggregates results from all oracle nodes (\textit{e.g.}, by taking the mean or median) and presents the final result to the consumer. Other DON solutions like WINkLink \cite{WINkLinkdoc} also utilize reputation mechanisms to promote the honest participation of oracle nodes.
Despite their popularity, existing DON solutions face data confidentiality and scalability challenges. The oracle service often requires the plaintext data be publicized and consumed on a smart contract in an open manner, which is not ideal for sensitive data items. At the same time, the data that a DON can feed to the blockchain are often limited in size, mainly due to the blockchain's intrinsic scalability limitation attributed by high on-chain costs for computation and storage \cite{xiao2023decentruth}. 
Each oracle node in a DON also needs to keep a list of ``premium sources'' and only collects data from them to maintain the quality of its data offering (\textit{i.e.}, the consumers) \cite{chainlink3levels}. Consequently, nodes in the DON tend to gradually converge on selecting from a limited list of well-known data sources. This tendency has posed another significant constraint on DONs' applicability to enabling data marketplaces.

\rev{DEXO shares similarities with DON solutions by relying on decentralized oracle nodes to mitigate single-point failures. However, traditional oracle services including DONs transmit data to smart contracts on the blockchain, which provides no data confidentiality and limits data volume and efficiency. In contrast, \sysname transmits data off-chain with on-chain settlement for exchange, achieving end-to-end confidentiality and greater scalability in data sizes. Additionally, while DONs rely on node operators to determine their own data sources, \sysname allows distributed data sources to participate in the data market directly by joining P-DApps. This shift reduces reliance on monopolistic data providers and fosters a more diverse data marketplace.
}









\section{System Model}
\label{sec:system-model}

\subsection{Participation Model}

We define four participant types in the \sysname data market:
\begin{itemize}
    \item \textbf{Data-providing DApp} (\textbf{P-DApp}, or \textbf{provider}) is a seller in the data market. It advertises the availability of certain \emph{off-chain} user-end data through \sysname and expects to receive compensation once \sysname facilitates a sale of the advertised data. \rev{A P-DApp is a DApp, comprising of a backend smart contract (called a \sysname contract) and a user-facing frontend server.}
    
    \item \textbf{P-DApp users} are P-DApp's end users, typically mobile or IoT device owners, who agree to participate in the P-DApp's data sales through \sysname. They are data sources of the P-DApp and will be compensated for data sales. \rev{P-DApp users can interact with P-DApp's frontend server through standard secure communication protocols such as TLS.}
    \item 
    \textbf{Consumer} 
    is a buyer in the data market. 
    A consumer can browse a P-DApp's \rev{data advertisements on \sysname and is willing to pay for an interested data item.}
    \item \textbf{\sysname node} is reminiscent of an oracle node in existing DON schemes (\textit{e.g.}, Chainlink). A fixed number of \sysname nodes constitute the \textbf{\sysname network} that jointly fulfills the data exchange mission between P-DApps and consumers. We assume there are fixed $N$ \sysname nodes in our system.
\end{itemize}

Besides the above roles, we assume that a \rev{\textbf{smart contract platform}}, such as Ethereum, is in place to serve as the backend environment of P-DApps and the \sysname network. \rev{We further assume a P-DApp prices its data items. It creates a smart contract that specifies the data description and predefined price prior to any exchanges with a consumer through \sysname. \sysname acts as a neutral platform that facilitates the exchanges and does not intervene in the pricing process. We leave more complex pricing schemes, such as auctioning, to future work.}

\begin{table}
    \centering
    \small
    \caption{List of Notations}
    \resizebox{0.48\textwidth}{!}{
    \begin{tabular}{cc}
        \toprule
        Notation & Description \\
        \midrule
        $N$ & Number of \sysname nodes \\
        $\mathcal{N}_i$ & \sysname node $i$ ($i\in[N]$) \\
        $\mathcal{C}_{\sysname}$ & \sysname contract \\
        $F$ & Maximum compromised \sysname nodes ($F<\frac{1}{2} N$) \\
        $t$ & Secret sharing threshold ($F< t\leq N-F$) \\
        $\mathcal{F}_{TA}$ & Trusted application within a TEE \\
        $\mathcal{F}_{sc}$ & Smart contract ideal functionality \\
        $\mathcal{G}_{att}$ & TEE attested execution functionality \\
        $\mathcal{F}_{ss}$ & Secrete sharing functionality \\
        $\mathcal{F}_{com}$ & Commitment ideal functionality \\
        $D_i$ & Raw data gathered by P-DApp user $i$\\
        $d_i$ & Pre-preprocessed and formatted data for trading \\
        $ds_{i,j}$ & Secret data share of user $i$ for \sysname node $j$ \\
        $\sigma_{i,j}$ & $\mathcal{F}_{TA}$'s signature over and $ds_{i,j}$ and TEE runtime \\
        \rev{$cid$} & \rev{Contract identifier for data exchange} \\
        \rev{$eid$} & \rev{TEE instance identifier} \\
        \rev{$mpk$} & \rev{Master public key for TEE attestation} \\
        \rev{$msk$} & \rev{Master secret key for TEE attestation} \\
        \rev{$\Delta_j$} & \rev{Merkle tree root hash of encrypted shares submitted by node $j$} \\
        \rev{$z$} & \rev{Encrypted data shares sent to consumers} \\
        \rev{$k_j$} & \rev{Secret key used by node $j$ to encrypt data shares} \\
        \bottomrule
    \end{tabular}
    }
    \label{tab:notation}
\end{table}

\subsection{Design Goals}
\label{subsec:goals}

DEXO aims to enable decentralized data exchange with the following objectives:


\textbf{O1: Data source verifiability.~}
To ensure the quality of the data collected from sources, \sysname requires each P-DApp user to \rev{pre-process locally gathered data using a provided function}. This \rev{pre-processing function ensures the data for sale conforms to a certain format and normality as advertised by the P-DApp and} should be verifiable for integrity by the \sysname network. 
\rev{
In this work, we do not require \sysname to provide broader data-derived quality control, such as assessing the contextual utility of the data. We leave such considerations to data consumers who can decide to purchase future data from the P-DApp.}

\textbf{O2: End-to-end data confidentiality.~}
The data requested by a consumer should only be revealed to the consumer; it remains confidential to the \sysname nodes and the public.

\textbf{O3: Fault tolerance and no single-point failure.~} 
The \sysname network is decentralized, with multiple nodes working to facilitate the data verification and exchange process. The above objectives should still be accomplished when a minority of nodes are compromised and do not follow the correct protocol. 

\textbf{O4: Fair exchange.~} \sysname should facilitate a fair exchange between a P-DApp and a consumer when the latter requests the former's data. The P-DApp should receive compensation only if the consumer obtains the correct data, and vice versa. 

\textbf{O5: Off-chain data delivery.~}
\sysname should facilitate a fast exchange process for off-chain data delivery and on-chain verifiability. The blockchain costs should be minimized. This is a key difference from the existing DON schemes where the entirety of requested data has to be delivered on-chain, constraining the data size due to the on-chain cost.

In particular, realizing the objectives requires handling faulty behaviors and potential collusion among the participants. The detailed threat model is provided in Section \ref{subsec:threat-model}.

\subsection{Threat Model}
\label{subsec:threat-model}

Out of the $N$ nodes in the \sysname network, we assume at most $F<\frac{1}{2}N$ of them are compromised at any point and the rest will operate correctly. This threshold assumption is sufficient to encompass that of existing oracle networks such as Chainlink \cite{chainlink3levels} where at most 6 out of the 21 oracle nodes could be compromised.
Specifically, the malicious activities of a compromised node relevant to \sysname's operation include:
\begin{enumerate}
    \item not following the designated protocol and sending arbitrary information to other participants in the system;
    \item extracting a P-DApp user's data shares and exposing them in untrusted domains;
    \item colluding with a P-DApp to scam a consumer for payment without providing the requested data in full;
    \item colluding with a consumer to scam a P-DApp for any portion of useful data without a full payment.
\end{enumerate}
The last two colluding situations imply that the P-DApp or consumer may not execute a given data exchange protocol faithfully. In light of this, \sysname should be able to allow either party to abort the exchange safely.

For the data source aspect, we assume the P-DApp users are responsible for managing their own raw data source. They will be ultimately compensated for providing high-quality data. We require that a user's locally established TEE is tamper-proof and performs attested execution of certain data pre-processing rules required by the \sysname system. 
Specifically, \rev{The TEE safeguards the pre-processed data by isolating it within a secure enclave, preventing unauthorized access or tampering even from the host operating system. The TEE attestation provides proof of TEE program's integrity and TEE hardware's authenticity. }
Furthermore, the trusted application $\mathcal{F}_{TA}$ in TEE is ratified by the \sysname community and available in the public domain. It should always correctly execute the data pre-processing and security functions (\textit{e.g.}, secret sharing and generating attestation reports) following the standard TEE security properties. We also assume the P-DApp server is trusted by its users for not leaking their data before the sale. Lastly, the TEE attestation service is trusted for verifying the TEE platform's authenticity when it receives an attestation report forwarded by a \sysname node.




\section{Building Blocks}
\label{sec:prelim}


In this section, we describe the building blocks of \sysname, including their key properties and ideal functionalities.
Using the ideal functionalities allows us to abstract away their implementation and focus on composing the \sysname system.

\label{sec:preliminaries}

\subsection{Smart Contract}

Smart contracts enable the automatic and traceable execution of multiparty business logic and typically live in an append-only blockchain ledger. We adopt the
ideal blockchain functionality $\mathcal{F}_{blockchain}$ proposed in \cite{ekiden} as the baseline. It allows participants to read on-chain information through the $\mathbf{read}()$ interface and append new information through the $\mathbf{write}()$ interface. 
Here we describe an ideal smart contract functionality $\mathcal{F}_{sc}$ by extending $\mathcal{F}_{blockchain}$ to incorporate more expressive contract operations as follows:


\vspace{5pt}
\noindent\fbox{\begin{minipage}{.472\textwidth}
\begin{definition}[Ideal Functionality $\mathcal{F}_{sc}$] The ideal smart contract functionality $\mathcal{F}_{sc}$ inherits $\mathcal{F}_{blockchain}$'s persistent storage $\mathrm{LStorage}$ and supports the following interfaces:
\begin{itemize}
    \item $\mathcal{F}_{sc}.\mathbf{create}(\mathcal{C},params[])$ creates a smart contract with a given contract encoding $\mathcal{C}$ and initializing parameters $params[]$. If successful, it generates a contract identifier $cid$, writes the contract object into $\mathrm{LStorage}$, and returns \textsc{success} and $cid$ to the sender.
    \item $\mathcal{F}_{sc}.\mathbf{read}(cid,\textit{``var''})$ looks up the smart contract identified by $cid$ and the state variable identified by \textit{var} within $\mathrm{LStorage}$. If \textit{var} exists, it returns the variable value to the sender.
    \item $\mathcal{F}_{sc}.\mathbf{write}(cid,\textit{``func''},args[])$ writes to the smart contract identified by $cid$ invoking the specified function identified by \textit{func} with arguments $args[]$. If successful, it updates the modified state variables in $\mathrm{LStorage}$ and returns \textsc{success} to the sender.
\end{itemize}
\end{definition}

\end{minipage}}

\subsection{TEE-based Attested Execution}
\label{subsec:buildingblock-tee}
TEE is a secure area within a processor that provides an isolated and protected environment for executing sensitive code and handling confidential data. 
TEEs are designed to ensure the confidentiality, integrity, and authenticity of the data and code running within them even amid malicious software or hardware attacks on the hosting system \cite{costan2016intel}. 
Popular TEE platforms include Intel Software Guard eXtensions (SGX)~\cite{sgx:white1}, 
AMD Secure Encrypted Virtualization (SEV)~\cite{amdsecure}, ARM TrustZone~\cite{ARMTrustZone},
and Apple Secure Enclave in T2 chip~\cite{apple2019T2}, showing a diverse ecosystem of TEE implementations on various CPU architectures. 
A key functionality of TEE is attested execution that safely executes the TEE program while proving the program's authenticity and integrity. A signature for the enclave is created by using a hard-coded key based on the TEE initial state, code, and data and then verified with the help of chip vendors \cite{Intel-Attestation-1,OP-TEE-Attestation-Concept1}. 
In this paper, we assume that TEE capability is available for mobile and IoT devices and adopt the generalized attested execution functionality $\mathcal{G}_{att}$ defined in \cite{pass2017formal} (while ARM TrustZone is used for experiments). Here we provide a simplified description of $\mathcal{G}_{att}$:



\vspace{5pt}
\noindent\fbox{\begin{minipage}{.472\textwidth}
\begin{definition}[Ideal Functionality $\mathcal{G}_{att}$]
$\mathcal{G}_{att}$ is the ideal functionality of general TEE-based attested execution. It is hard-coded with a public-private key pair $(mpk,msk)$, keeps a persistent TEE memory $\mathrm{TMem}$, and provides the following interfaces:
\begin{itemize}
    \item $\mathcal{G}_{att}.\mathbf{install}(\mathcal{F}_{TA})$ establishes a new TEE instance inside $\mathrm{TMem}$ from the the caller-provided trusted application $\mathcal{F}_{TA}$. $\mathcal{F}_{TA}$'s state variables are also stored in $\mathrm{TMem}$. If successful, it generates an identifier $eid$ for the new TEE instance and returns $eid$ to the caller.
    \item $\mathcal{G}_{att}.\mathbf{resume}(eid,args[])$ executes the program (\textit{i.e.} $\mathcal{F}_{TA}$) inside the $eid$-identified TEE instance with the given arguments $args[]$. If successful, it returns the execution result $res$ and a signature over the TEE runtime $\sigma^{rt}$ signed by $msk$. Any modified state variables are updated in $\mathrm{TMem}$.
\end{itemize}
\end{definition}
\end{minipage}}
\vspace{1pt}

The $\mathcal{G}_{att}.\mathbf{resume}()$ essentially fulfills the attested execution functionality with $\sigma^{rt}$ attesting to the authenticity and integrity of the TEE program $\mathcal{F}_{TA}$. 
\rev{TEE attestation ensures that all data processed by the TA remains confidential and tamper-proof. The TEE safeguards the data by isolating it within a secure enclave, preventing unauthorized access even from the host operating system or hardware. The attestation mechanism generates a signed report to verify that the data was handled securely by the TA, ensuring that no unauthorized modifications occurred during processing.}
We will describe a detailed $\mathcal{F}_{TA}$ and the $\mathbf{resume}$ procedure in Section \ref{subsec:design-user-routine}.

\subsection{Shamir's Secret Sharing} 

Secret sharing is a cryptographic technique to distribute a confidential datum $d$ into multiple ($n$) fragments, known as shares. A $(t,n)$-secret sharing scheme ensures that anyone with no fewer than $t$ of the shares can reconstruct $d$.  
 for which we describe the ideal functionality as follows: 

\vspace{5pt}
\noindent\fbox{\begin{minipage}{.472\textwidth}
\begin{definition}[Ideal Functionality $\mathcal{F}_{ss}$]
The ideal secret sharing functionality $\mathcal{F}_{ss}$ provides two interfaces:
\begin{itemize}
    \item $\mathcal{F}_{ss}.\mathbf{createshares}(t,n,d)$ generates $n$ shares from the provided secret $d$ so that any $t$ out of the shares can be used to reconstruct $d$. 
    \item $\mathcal{F}_{ss}.\mathbf{reconstruct}(t,n,ss[])$ either returns the secret $d$ successfully reconstructed from the set of shares $ss[]$ or an error indicating there are inconsistencies in $ss[]$ that preventing the reconstruction of a single secret.
\end{itemize}
\end{definition}
\end{minipage}}
\vspace{1pt}

In \sysname, Shamir's Secret Sharing \cite{shamir1979share} is used to achieve end-to-end data confidentiality and resist single-point failures. We disperse the data into multiple shares which are provided to different \sysname nodes. It ensures any smaller subset of shares below this threshold remains oblivious to any useful knowledge of the original data 
so that even if a subset of nodes falls prey to security breaches, the integrity and confidentiality of the data remain untarnished.

\subsection{Cryptographic Commitment}

A cryptographic commitment allows one to commit to a message while keeping it hidden from other parties, with the ability to open the committed message later \cite{Goldreich_2001}. It crucially achieves the binding property---the committing party cannot claim a different message was committed.
We adopt the general single-message commitment functionality $\mathcal{F}_{com}$ defined in \cite{canetti2001uccommitments} and provide a concise description as follows:

\vspace{5pt}
\noindent\fbox{\begin{minipage}{.472\textwidth}
\begin{definition}[Ideal Functionality $\mathcal{F}_{com}$] The ideal 
single-message commitment functionality $\mathcal{F}_{com}$ keeps a persistent commitment storage $CStorage$ provides the following interfaces:
\begin{itemize}
    \item $\mathcal{F}_{com}.\mathbf{commit}(sid,P_i,P_j,msg)$ allows the caller $P_i$ to generate a commitment generates a commitment $com$ to $msg$ of sequence number $sid$. If successful, it stores $(sid,com)$ in $CStorage$, forwards $msg$ to $P_j$ (if $P_j$ is specified), and return \textsc{success} to $P_i$.
    \item $\mathcal{F}_{com}.\mathbf{open}(sid,P_i,P_j)$ allows the caller $P_i$ to open its previous commitment identified by $sid$ from $CStorage$ and discloses the original $msg$ to $P_j$ (if $P_j$ is specified).
\end{itemize} 
\end{definition}
\end{minipage}}
\vspace{1pt}

The commitment scheme has been used for constructing fair exchange protocols \cite{dziembowski2018fairswap,eckey2020optiswap}. In \sysname, we employ a similar construction of fair exchange protocol as in OptiSwap \cite{eckey2020optiswap} with a modification that a \sysname node commits to each data secret share rather than the original data.

\section{\sysname Design}
\label{sec:detailed-design}


\subsection{System Overview}

\rev{Achieving the objectives outlined for DEXO requires addressing several unique design challenges that existing DONs or decentralized data exchange solutions fail to tackle. We integrate TEE-based secret sharing with smart contract mechanisms to enable end-to-end data confidentiality while maintaining data verifiability. This integration ensures secure handling of sensitive data without exposing plaintext to any intermediary, which commercial DONs do not achieve. We also design protocols to achieve fault tolerance in the exchange process, allowing the system to handle misbehavior by up to \( F \) compromised DEXO nodes while still delivering data successfully to consumers. In contrast, existing decentralized data exchange solutions rely on trusted intermediaries or single-point designs that remain vulnerable to such faults.}

\begin{figure*}
    \centering
    \includegraphics[width=.95\linewidth]{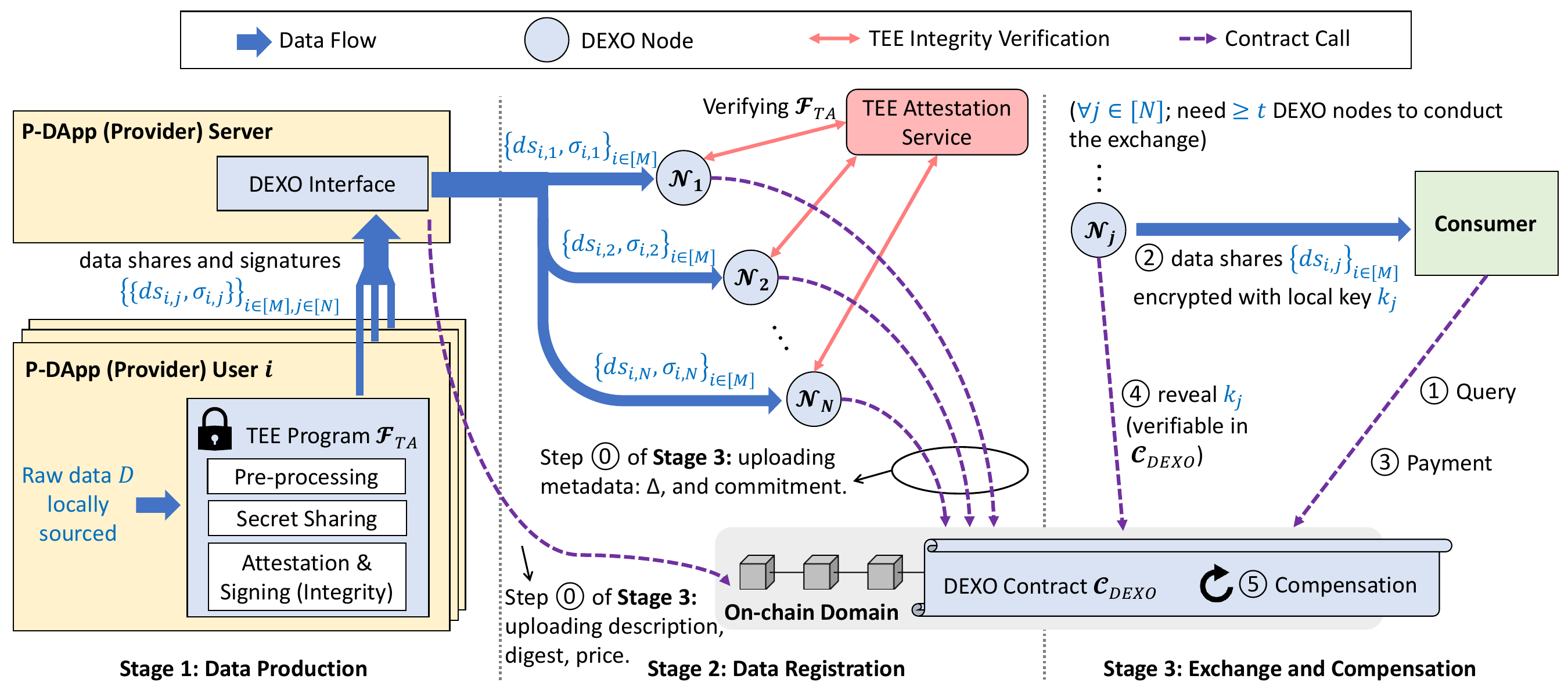}
    \caption{DEXO System Architecture and Workflow}
    \label{fig:arch}
\end{figure*}

Now we describe \sysname's architectural design and high-level workflow, as shown in Fig. \ref{fig:arch}. To simplify the description, we consider the case of one P-DApp and one consumer. We assume the P-DApp has $M$ end users who can contribute local data to the P-DApp's data sale. We assume there are $N$ \sysname nodes that are pre-determined, denoted $\mathcal{N}_1,...,\mathcal{N}_N$.

\vspace{2pt}
\noindent
\textbf{Stage 0: Initialization}\rev{---This stage aims to set up the foundational components for secure data exchange, including configuring the P-DApp server, enabling TEE functionality on user devices, and deploying the DEXO Contract for fair exchanges.} 

To become a seller (\textit{i.e.}, P-DApp) in \sysname's data exchange market, a DApp needs to make certain architectural modifications on both its frontend server and user ends. The P-DApp server can establish TLS connections with all \sysname nodes to forward user data to the \sysname network. We assume each user can establish a secure TLS connection with the P-DApp server as in most web applications.

Each user who is willing to participate in the data offering needs to enable the TEE functionality on their device. They establish a new TEE container to instantiate a \textbf{trusted application} $\mathcal{F}_{TA}$ which is ratified by \sysname and available in the public domain. ${\mathcal{F}_{TA}}$ performs data processing and generation of data shares in Stage 1 as we will describe shortly. Once $\mathcal{F}_{TA}$ is instantiated in a TEE container, its program integrity can be verified by the server and any \sysname node through remote attestation.

The P-DApp server creates a dedicated \textbf{\sysname Contract} $\mathcal{C}_{\sysname}$ that follows a pre-defined format (see \S\ref{subsec:dexo-contract}) and designates the \sysname notes as curators. 
$\mathcal{C}_{DEXO}$ acts as an adjudicator of fair exchange, holding the payment and ensuring its transfer to the P-DApp users if the exchange is completed or returning it to the consumer if the correct data is not received.
We assume the \sysname nodes have access to the attestation service for each type of TEE platform. Attestation services are normally provided by the TEE vendors and are generally assumed to behave honestly.

\vspace{2pt}
\noindent
\textbf{Stage 1: Data Production}\rev{---This stage aims to process raw data from P-DApp users, generate secure and verifiable data shares using TEE, and initialize a smart contract with the data description and pricing information for exchange.} This stage involves the server and users of a P-DApp. 
A P-DApp user $i$ gathers raw data $D_i$ and feeds them into the TEE program $\mathcal{F}_{TA}$. $\mathcal{F}_{TA}$ performs the following tasks: 
\begin{itemize}
    \item \textbf{Pre-processing:} $D_i$ is processed per a given rule (e.g., maximum value range, moving average) and converted into a certain format. The formatted result is denoted $d_i$.
    \item \textbf{Secret Sharing:} $d_i$ is split into $N$ shares via a $(t,N)$-secret sharing algorithm, with each share denoted $ds_{i,j}$ for $j\in[N]$. One can reconstruct $d_i$ with at least $t$ shares. We require $F<t\leq N-F$ to ensure a safe reconstruction of $d_i$ by the consumer (to prove in Section \ref{sec:analyses}).
    \item \textbf{Signature Generation:} Following the data production, a signature $\sigma_{i,j}$ is created based on each share $ds_{i,j}$ and the TEE runtime measurement with the TEE platform private key. This signature proves the integrity of the data share generation and the authenticity of the TEE platform.
\end{itemize}

When $\mathcal{F}_{TA}$ finishes, the P-DApp Server creates a data item on the smart contract, initialized with a description $\{desc\}$, along with the price for that data. The \rev{P-DApp} server also sends addresses of data sources and DEXO nodes to the contract. This process also serves as Step \circled{0} for the exchange protocol in Stage 3 together with the ``Ready to Exchange'' step in Stage 2.

\vspace{2pt}
\noindent
\textbf{Stage 2: Data Registration}\rev{---This stage aims to verify the authenticity of data shares through attestation, prepare encrypted shares for exchange, and register the metadata and cryptographic commitments with the DEXO Contract.}

Upon receiving the $j^{th}$ shares and signatures from all $M$ users of the P-DApp, \textit{i.e.}, $\{ds_{i,j},\sigma_{i,j}\}$ for $i\in[M]$, \sysname node $\mathcal{N}_j$ performs the following steps:
\begin{itemize}
    \item \textbf{Source Attestation:} 
    $\mathcal{N}_j$ verifies the authenticity and integrity of data share $ds_{i,j}$ as well as user $i$'s TEE platform with the help of an attestation server. This step ensures $ds_{i,j}$ can be only accepted if it is produced by the required $\mathcal{F}_{TA}$ on an unaltered TEE platform. 
    \item \textbf{Ready to Exchange:} After source attestation, $\mathcal{N}_j$ needs to use data shares to prepare for the exchange transaction. $\mathcal{N}_j$ first generates a local secret key $k_j$ and encrypts the local data shares $\{ds_{i,j}\}_{i\in[M]}$ to get the ciphertext $z_j$. A data digest $\Delta_j$ is generated from $z_j$ and so is a cryptographic commitment $com_j$ on $k_j$. $\mathcal{N}_j$ then submits $\Delta_j$, $com_j$ to $\mathcal{C}_{\sysname}$ in one contract call. The $\{\Delta_j,com_j\}$ submitted by all $j\in[N]$ constitute the metadata for P-DApp.
    This process also serves as the Step \circled{0} for the exchange protocol in Stage 3.
\end{itemize}

\vspace{2pt}
\noindent
\textbf{Stage 3: Exchange and Compensation}\rev{---This stage aims to facilitate the fair exchange of data for payment, ensure data reconstruction integrity, handle potential disputes, and distribute compensation to P-DApp users.}
\begin{itemize}
    \item \textbf{Fair Exchange:} It is started by the consumer with a query to the $\mathcal{C}_{\sysname}$ based on the data description $desc$ (Step \circled{1}) \rev{and establish a TLS connection with each \sysname node.}
    Upon observing the query on-chain, node $\mathcal{N}_j$ ($\forall j\in[N]$) sends the ciphertext $z_j$ to the consumer (Step \circled{2}). Then the consumer verifies the integrity of $z_j$ with the corresponding on-chain metadata (data shares digest) before calling to $\mathcal{C}_{\sysname}$ to indicate acceptance with payment (Step \circled{3}). Then $\mathcal{N}_j$ calls $\mathcal{C}_{\sysname}$ to reveal its local key $k_j$, whose integrity can be automatically verified within $\mathcal{C}_{\sysname}$ (Step \circled{4}). Once verified, consumer can retrieve $k_j$ from $\mathcal{C}_{\sysname}$ and use it to decrypt $z_j$ to get $\{ds_{i,j}\}_{i\in[M]}$.
    \item \textbf{Data Reconstruction and (Optional) Dispute Handling.} Once decrypting the shares, the consumer can reconstruct the original data. If both parties conduct their operation with integrity, this stage is unnecessary for the buyer or seller. However, when the consumer finds that the shares cannot be used for reconstruction or the reconstructed data item does not conform to the promised attributes specified in the contract, the buyer can initiate a dispute. We provide more details of dispute handling in Section \ref{subsec:dexo-contract}.

    \item \textbf{Provider Compensation:} If all goes successfully (no dispute), $\mathcal{C}_{\sysname}$ distributes the consumer's payment to the P-DApp's data-providing users, fulfilling the final compensation (Step \circled{5}). 
\end{itemize}

In what remains of this section, we describe each participant routine in the $(\mathcal{G}_{att},\rev{\mathcal{F}_{sc}},\mathcal{F}_{ss},\mathcal{F}_{com})$-hybrid model based on the Universal Composability (UC) framework \cite{canetti2001universally}, utilizing the ideal functionalities described in Section \ref{sec:prelim}.

\subsection{P-DApp User Routine}
\label{subsec:design-user-routine}

As the distributed data sources for \sysname, P-DApp users need to execute the \sysname-ratified $\mathcal{F}_{TA}$ for pre-processing raw data and generating data shares. The P-DApp user routine is shown in Algorithm \ref{alg:user-tee}.
Once $\mathcal{F}_{TA}$ is installed on the user device $\mathcal{G}_{att}.\mathbf{install}(\mathcal{F}_{TA})$, the server can attest to $\mathcal{F}_{TA}$'s integrity by sending an \textsc{Attest} command and then collect data from the user by sending a \textsc{Solicit} command. Once solicited, user $i$ resumes to the $\mathcal{F}_{TA}$ at the \textsc{GenData} command, which produces the secret shares $\{ds_{i,j}\}_{j\in[N]}$ and corresponding signatures $\{\sigma_{i,j}\}_{j\in[N]}$. It is worth noting that $\sigma_{i,j}$ results from signing $ds_{i,j}$ and the runtime measurement with the TEE platform private key. The secret shares and signatures are sent back to the P-DApp server, which disseminates them to the different \sysname nodes accordingly. Each share's signature accompanies it as it moves through the system. \rev{To ensure the appropriate distribution of secret shares, the TEE program $\mathcal{F}_{TA}$ also specifies the destination DEXO node for each share during its generation, for instance, share $ds_{i,j}$ is destined to node $j$. The P-DApp server then relays the shares to the specified nodes. For secure data transmission, the P-DApp server can leverage standard secure communication mechanisms to deliver the data shares to \sysname nodes, such as establishing a TLS connection with each DEXO node which acts as a public web server.}


When the P-DApp server receives data from various data sources, it needs to prepare certain information to construct the smart contract. This information includes ${desc}$, price, and addresses of data sources and DEXO nodes.




\SetKwInput{KwParticipants}{Participants}
\SetKwInput{KwParameters}{Parameters}
\SetKwInput{KwInit}{Init}
\SetKwInput{KwRequestDataUse}{RequestDataUse}
\SetKwInput{KwRequest}{Request}
\SetKwInput{KwFreeze}{Freeze}
\SetKwInput{KwCompute}{Compute}
\SetKwInput{KwFinalize}{Finalize}
\SetKwInput{KwCancel}{Cancel}
\SetKwInput{KwRecord}{Record}
\SetKwInput{KwRevokeContract}{RevokeContract}
\SetKwInput{KwRevoke}{Revoke}
\SetKwInput{KwCompComplete}{ComputationComplete}
\SetKwInput{KwCompleteTrans}{CompleteTransaction}

\begin{algorithm}[!ht]
\SetKwProg{KwOnInit}{On init}{:}{}
\SetKwProg{KwOnReceive}{On receive}{:}{}
\SetKwProg{KwOnInput}{On input}{:}{}
\SetKwProg{KwOnSecretInput}{On secret input}{:}{}
\SetKwProg{KwInitFunc}{Initialize}{:}{}
\SetKwProg{KwOnResume}{On resume}{:}{}
\SetKwProg{KwEncrypt}{encrypt}{:}{}
\SetKwProg{KwCompleteTx}{completeTx}{:}{}
    \setstretch{1.1}
    \raggedright
	\caption{P-DApp User $i$ Routine}
	\label{alg:user-tee}
    \small


\KwParameters{$t,N$}

{\color{gray}\tcc{Normal Routine (insecure world)}}
\KwOnInit{}{
    $eid \gets\mathcal{G}_{att}.\mathbf{install}(\mathcal{F}_{TA})$;\\
}

\vspace{2pt}
\KwOnReceive{\textnormal{$(\textsc{``Attest"})$ from $Server$}}{
    $(rt\_msmt,\sigma^{rt})\gets\mathcal{G}_{att}.\mathbf{resume}(eid,\textsc{``Attest"})$;\\
    Send $(\textsc{``AttReport"},rt\_msmt,\sigma^{rt})$ to $Server$;
}

\vspace{2pt}
\KwOnReceive{\textnormal{$(\textsc{``Solicit"})$ from $Server$}}{
    Gather rawdata $D$;\\
    $(\{ds_i\},\{\sigma_i\},{mpk_i})\gets\mathcal{G}_{att}.\mathbf{resume}(eid,\textsc{``GenData"},N,D)$;\\
    $\forall j\in[N]$: send $(\textsc{``DataShare"},j,ds_j,\sigma_j,mpk_i)$ to $Server$;
}

\vspace{2pt}
{\color{gray}\tcc{Trusted Application $\mathcal{F}_{TA}$ (the TEE program, executed upon $\mathcal{G}_{att}.\mathbf{resume}()$)}}
\KwOnResume{\textnormal{$(eid,args[])$}}{
    Generate TEE runtime measurement $rt\_msmt$; \\
    Retrieve TEE platform public-private key pair $(mpk,msk)$; \\
    $\sigma^{rt}\gets \mathbf{Sign}_{msk}(rt\_msmt)$; {~\color{gray}// signature of runtime env.} \\
    \If{$(\textsc{``Attest"}\in args[])$}{
        \textbf{return} $(rt\_msmt,\sigma^{rt},mpk)$;\\
    }
    \If{$(\textsc{``GenData"}\in args[])$}
    {
        Read $N,D$ from $args[]$;\\
        Pre-process $D$ and convert it to the required format, get $d$;\\ 
        $(ds_1,ds_2,...,ds_N)\gets\mathcal{F}_{ss}.\mathbf{createshares}(t,N,d)$;\\
        $\sigma_j\gets\mathbf{Sign}_{msk}(ds_j,rt\_msmt)$;\\
        \textbf{return} $(\{ds_j\}_{j\in[N]},\{\sigma_j\}_{j\in[N]},mpk)$; \\
    }
}



\setstretch{1}
\end{algorithm}

\subsection{\sysname Node Routine}
\label{subsec:design-node-routine}


\sysname incorporates attestation and verification mechanisms to establish trust in user-provided data. As is shown in Algorithm \ref{alg.seller_routing}, 
\sysname nodes bear the responsibility of managing the participation of DApps in the system and verifying the integrity of data shares provided by each P-DApp user. 

\SetKwInput{KwParameters}{Parameters}
\SetKwInput{KwInit}{Init}
\SetKwInput{KwRequestDataUse}{RequestDataUse}
\SetKwInput{KwRequest}{Request}
\SetKwInput{KwFreeze}{Freeze}
\SetKwInput{KwCompute}{Compute}
\SetKwInput{KwFinalize}{Finalize}
\SetKwInput{KwCancel}{Cancel}
\SetKwInput{KwRecord}{Record}
\SetKwInput{KwRevokeContract}{RevokeContract}
\SetKwInput{KwRevoke}{Revoke}
\SetKwInput{KwCompComplete}{ComputationComplete}
\SetKwInput{KwCompleteTrans}{CompleteTransaction}

\begin{algorithm}[!ht]
\small
\raggedright
\caption{\sysname Node Routine}
\label{alg.seller_routing}
\setstretch{1.1}

\SetKwProg{KwOnReceive}{On receive}{:}{}
\SetKwProg{KwOnInput}{On input}{:}{}

\vspace{2pt}
{\color{gray}\tcc{Data source verification}}
\KwOnReceive{\textnormal{$(\textsc{``DataShares"},\{ds\},Signatures,pubkeys,desc,userIDs)$ from a P-DApp $Server$}}{
    Send $(\textsc{``VerifyTEE"},tid,pubkeys)$ to $Attestation~Server$; \\
    Store \rev{$DS[tid]\leftarrow$} $(\{ds\},desc,userIDs)$; 
}

\vspace{2pt}
{\color{gray}\tcc{Data Publication}}
\KwOnReceive{\textnormal{$(\textsc{``TEEverified"},tid)$ from $Attestation~Server$}}{
    Generate secret key $k$;\\
    $x[]\leftarrow \rev{DS[tid]}$;\\
    $z[]\leftarrow \mathbf{Encrypt}_k(x[])$;\\
    $\Delta\leftarrow \mathbf{MTHash}(z[])$; {\color{gray}//Merkle tree root hash}\\
    $com\leftarrow \mathcal{F}_{com}.\mathbf{commit}(tid,self,\_,k)$;\\
    $\rev{\mathcal{F}_{sc}}.\mathbf{write}(cid,``initialize", \rev{tid}, seller, price, desc, \Delta, com)$; 
}

\vspace{2pt}
{\color{gray}\tcc{If a buyer has queried the data}}
\KwOnReceive{\textnormal{$(\textsc{``NoticeBuy"},cid,BuyerID))$ from $Buyer$}}{
    $status\gets\rev{\mathcal{F}_{sc}}.\mathbf{read}(cid,``buyerStatus\rev{.BuyerID}")$; \\
    If $status=\textsc{queried}$: Send $z[]$ to $Buyer$; \\
}

\vspace{2pt}
{\color{gray}\tcc{If $z[]$ is accepted}}
\KwOnReceive{\textnormal{$(\textsc{``NoticeAccept"},cid,BuyerID))$ from $Buyer$}}{
    $status\gets\rev{\mathcal{F}_{sc}}.\mathbf{read}(cid,``buyerStatus\rev{.BuyerID}")$; \\
    If $status=\textsc{accepted}$: 
    $\rev{\mathcal{F}_{sc}}.\mathbf{write}(cid,``revealKey'',k)$;\\
}

\setstretch{1}
\end{algorithm}

Specifically, node $\mathcal{N}_j$ shall receive data shares $\{ds_{i,j}\}_{i\in[M]}$ and signatures $\{\sigma_{i,j}\}_{i\in[M]}$ if the P-DApp has $M$ users to provide data. For user $i$'s data, $\mathcal{N}_j$ needs to verify the data generation with the help of an attestation server. Upon receiving $\sigma_{i,j}$ and $mpk_i$ of user $i$, the attestation server can verify the authenticity of user $i$'s TEE platform (\textit{i.e.}, whether it is a genuine TEE hardware) by checking against the public key certificate for $mpk_i$. Once validated, the attestation server notifies $\mathcal{N}_j$ of the success and the latter can then use $mpk_i$ to validate the signature $\sigma_{i,j}$. If validated, $\mathcal{N}_j$ knows that the data share $ds_{i,j}$ has not been tampered with and can safely proceed to publicize $\{ds_{i,j}\}_{i\in[M]}$ onto $\mathcal{C}_{\sysname}$. 
\rev{$\mathcal{N}_j$ then encrypts $\{ds_{i,j}\}_{i\in[M]}$ with a newly generated secret key $k_j$, resulting in ciphertext $z_j$. It also computes a cryptographic digest of $z_j$, denoted $\Delta_j$. It can be conveniently fulfilled by the Merkle tree root ($\mathbf{MTHash}$) of all fixed-size fragments of $z_j$. The key $k_j$ is fed to the commitment scheme $\mathcal{F}_{com}.\mathbf{commit}$ so that the result $com_j$ can later be opened to verify the integrity of $k_j$. It performs step \circled{0} of the data exchange, by uploading $com_j,\Delta_j$ to the $\mathcal{C}_{\sysname}$.} 
At this point, node $\mathcal{N}_j$ is ready to process data queries from consumers.

Once a consumer indicates an interest in the P-DApp's data (step \circled{1}), $\mathcal{N}_j$ will need to provide its encrypted data shares $z_j$ to the consumer (step \circled{2}). Upon receiving an initial payment from the consumer to $\mathcal{C}_{\sysname}$ (\circled{3}), $\mathcal{N}_j$ then discloses the secret key $k_j$ in plaintext to $\mathcal{C}_{\sysname}$, where $k_j$ will be validated by opening the previous commitment $com$ on the  (\circled{4}). A success will start a count-down from a $timeout$ value (included in $desc$) which provides a buffer for the consumer to submit any dispute. When the timeout passes, the payment will be automatically redistributed to the P-DApp users as compensation.
Interactive dispute handling is described in Section \ref{subsec:dexo-contract}.

\subsection{Consumer Routine}
\label{subsec:consumer-routine}

The consumer routine is shown in Algorithm \ref{alg.buyer_routing}. It requires the consumer to actively observe existing \sysname contracts for any data of interest. Once determined \rev{the target contract $\mathcal{C}_{\sysname}$ and data item indexed by $dataID$}, \rev{the consumer retrieves the corresponding} $desc_j$ and the data shares digest $\Delta_j$ for all possible $j\in[N]$. \rev{To declare their intent, the consumer submits a query to $\mathcal{C}_{\sysname}$ using their cryptocurrency account address} (step \circled{1}). \rev{This wallet address is recorded in the contract for later payment processing and ensures secure identification throughout the exchange. The query can also provide optional user identifiers, such as IP address or email, for future auditing purposes.}
Then the consumer connects to at least $t$ \sysname nodes (we use $\mathcal{N}_{[t]}$ to denote the set of their indices) and requests for their encrypted data. \rev{This requires the consumer to establish a Web connection (usually over TLS) as a client with each \sysname node and pass authentication for the ownership of its cryptocurrency account address. This process is similar to the interaction between existing Web3 frontend servers and Web3 users, where third-party account management software (such as MetaMask and Alchemy) can be integrated into the Web interface to facilitate account authentication. } 

Once receiving $z_j$ from $\mathcal{N}_j$, the consumer verifies its integrity against $\Delta_j$, by recomputing the Merkle hash root. If it is verified, the consumer calls $\mathcal{C}_{\sysname}$'s $accept$ function with the required payment (step \circled{3}). The payment is now temporarily held in $\mathcal{C}_{\sysname}$'s balance. Once $\mathcal{N}_j$ releases $k_j$ to $\mathcal{C}_{\sysname}$ and the verification with commitment passes (step \circled{4}), the consumer can safely obtain $k_j$ to decrypt the corresponding $z_j$ to get $\{ds_{i,j}\}_{i\in[M]})$. When $t$ keys are obtained, the consumer can finally reconstruct the original data $d_i$ from $\{ds_{i,j}\}_{j\in\mathcal{N}_{[t]}})$ by using $\mathcal{F}_{ss}.\mathbf{reconstruct}(t,n,\{ds_{i,j}\}_{j\in\mathcal{N}_{[t]}})$.



\subsection{Fair Exchange with DEXO Contract}
\label{subsec:dexo-contract}

The previous sections have described the basic process of a fair exchange protocol session between a \sysname node and a consumer. In \sysname,
we employ a parallel composition of OptiSwap \cite{eckey2020optiswap} for a fair data exchange process between a consumer and the \sysname oracle nodes (each of who has a secret share of the consumer's requested data) with the help of the \sysname contract $\mathcal{C}_{DEXO}$ (see Algorithm \ref{alg:contract}). All parties are aware of a predicate function $\phi()$ that involves secret reconstruction to determine the misbehavior of individual \sysname nodes.
First, $\mathcal{C}_{DEXO}$ serves as the public storage of a P-DApp's metadata, \textit{i.e.}, $\{desc_j,\Delta_j,com_j,\}_{j\in[N]}$. This allows the consumer to verify the ciphertext data from the \sysname nodes (Step \circled{2}) and also serves as the temporary custodian of the payment (Step \circled{3}). It allows the opening of the commitment $com_j$ in one contract call (Step \circled{4}). If the opening succeeds, the $j^{th}$ portion of consumer payment is considered finalized.

\SetKwInput{KwParameters}{Parameters}
\SetKwInput{KwInit}{Init}
\SetKwInput{KwRequestDataUse}{RequestDataUse}
\SetKwInput{KwRequest}{Request}
\SetKwInput{KwFreeze}{Freeze}
\SetKwInput{KwCompute}{Compute}
\SetKwInput{KwFinalize}{Finalize}
\SetKwInput{KwCancel}{Cancel}
\SetKwInput{KwRecord}{Record}
\SetKwInput{KwRevokeContract}{RevokeContract}
\SetKwInput{KwRevoke}{Revoke}
\SetKwInput{KwCompComplete}{ComputationComplete}
\SetKwInput{KwCompleteTrans}{CompleteTransaction}

\begin{algorithm}[!ht]
\raggedright
\caption{Data Consumer Routine}
\label{alg.buyer_routing}
\small
\setstretch{1.1}

\SetKwProg{KwOnReceive}{On receive}{:}{}
\SetKwProg{KwOnInput}{On input}{:}{}


\vspace{2pt}
{\color{gray}\tcc{If a new transaction is found}}
\KwOnReceive{\textnormal{$(\textsc{``NoticeNewData"},cid,dataID)$ from $Self$}}{
    $nodeID,price,auxInfo,\Delta,dataSources)$;\\
    $(nodeID,price,auxInfo,\Delta)\gets\rev{\mathcal{F}_{sc}}.\mathbf{read}(cid,``dataSources\rev{.dataID}")$; \\
    If satisfied: $\rev{\mathcal{F}_{sc}}.\mathbf{write}(cid,``query'',dataID,buyerID)$; \\
}

\vspace{2pt}
{\color{gray}\tcc{Accept encrypted shares from selling node}}
\KwOnReceive{\textnormal{$(\textsc{``EncryptedShares"},z[],cid,dataID)$ from $\mathcal{N}_j$}}{
    Recompute $\Delta' \gets \mathbf{MTHash}(z[])$;\\
    $(nodeID,price,auxInfo,\Delta)\gets\rev{\mathcal{F}_{sc}}.\mathbf{read}(cid,``dataSources\rev{.dataID}")$; \\
    If $\Delta=\Delta'$: $\rev{\mathcal{F}_{sc}}.\mathbf{write}(cid,``accept'',price,buyerID)$; \\
}

\vspace{2pt}
{\color{gray}\tcc{Accept key}}
\KwOnReceive{\textnormal{$(\textsc{``NoticeKey"},cid,dataID)$ from $\mathcal{N}_j$}}{
    $k\gets\rev{\mathcal{F}_{sc}}.\mathbf{read}(cid,``keyRevealed\rev{.dataID}")$; \\
    Decryption: $x[]\leftarrow \mathbf{Decrypt}_k(z[])$;\\
    $\{ds_{i,j}\}_{i\in[M]}\leftarrow x[]$;\\
}

\vspace{2pt}
{\color{gray}\tcc{Once more than $t$ keys are obtained from \sysname nodes (node set: $\mathcal{N}_{[t]}$)}}
\KwOnReceive{\textnormal{$(\textsc{``Reconstruct"})$ from $Self$}}{
    Get $t,N$ and $timeout$ from the $auxInfo$ obtained from the previous $dataSources$ reading;\\
    For $i\in[M]$: $d_i\gets\mathcal{F}_{ss}.\mathbf{reconstruct}(t,n,\{ds_{i,j}\}_{j\in\mathcal{N}_{[t]}})$;\\
}

\vspace{2pt}
\rev{
{\color{gray}\tcc{Optional dispute procedure}}
\KwOnReceive{\textnormal{$(\textsc{``Dispute"})$ from $Self$}}{
{\color{gray}\tcc{Case 1: Reconstruct Original Data and Validate Against Description}}  
Initialize $valid = False$;\\
Select initial sets $S_1, S_2$ with $t+1$ shares each;\\
While $valid == False$: \\
\Indp
$d_1 \gets \mathcal{F}_{ss}.\mathbf{reconstruct}(t, n, S_1)$;\\
$d_2 \gets \mathcal{F}_{ss}.\mathbf{reconstruct}(t, n, S_2)$;\\
If $d_1 == d_2$: \\
\Indp
$valid = True$; {\color{gray}// Data consistency confirmed}\\
Set $d_{orig} = d_1$; {\color{gray}// Accept reconstructed data}\\
\Indm
Else: \\
\Indp
$S_1 \gets$ next combination of $t+1$ shares;\\
$S_2 \gets$ next combination of $t+1$ shares with at least one differing share;\\
\Indm
\Indm
Compare $d_{orig}$ with $desc$ from contract: \\
If $d_{orig} \neq desc$: \\
\Indp
$\rev{\mathcal{F}_{sc}}.\mathbf{write}(cid,``Challenge'',1,S_1,S_2)$;\\
\Indm
Else: Proceed without dispute.\\
{\color{gray}\tcc{Case 2: Detect and Verify Bad Shares After Reconstructing Data}}  
For each suspected bad share $ds_{i,j}$: \\
\Indp
Combine $ds_{i,j}$ with $t-1$ valid shares to reconstruct: \\
$d' \gets \mathcal{F}_{ss}.\mathbf{reconstruct}(t, n, S')$;\\
If $d' \neq d_{orig}$: \\
\Indp
$\rev{\mathcal{F}_{sc}}.\mathbf{write}(cid,``Challenge'',2,badshares)$;\\
\Indm
\Indm
}
}

\rev{
\KwOnReceive{\textnormal{$(\textsc{``noComplain"})$ from $Self$}}{
        $\rev{\mathcal{F}_{sc}}.\mathbf{write}(cid,``noComplain'',buyerID)$;\\
}
}

\setstretch{1}
\end{algorithm}

\textbf{Dispute handling.} 
\rev{
$\mathcal{C}_{DEXO}$ provides a dispute-handling interface (the $Challenge$ function) to protect consumers from inconsistencies in data received from \sysname. This ensures robust consumer protection and enhances the fair exchange process. We first define two key utilities used in verification:  
\begin{itemize}
    \item $\phi_1(t, n, shares[])$ verifies whether $shares[]$ can reconstruct the original data in the $(t,n)$-secret sharing scheme. $\phi_1()$ either returns a the reconstructed data $d$ or returns an error code.
    \item $\phi_2(d,desc)$ verifies if the data $d$ conforms to the format and normality description in $desc$ (e.g., whether $d$ is within the numerical range). If verifies, $\phi_2()$ returns True; otherwise False.
\end{itemize}
$\phi_1()$ and $\phi_2()$ are implemented as $\mathcal{C}_{DEXO}$'s internal functions (implementation details are obviated in Algorithm \ref{alg:contract}).
}


\rev{Specifically, $\mathcal{C}_{DEXO}$ supports the following two dispute scenarios.  
\begin{enumerate}  
    \item \textbf{Reconstructed data does not follow description:}  
    If the consumer reconstructs the dataset $\{d_i\}_{i\in[M]}$ from received shares, but it does not match the format or attributes described in the contract, the consumer can invoke the contract's $challenge$ function. The contract first verifies share validity using Merkle Tree proofs, then applies $\phi_1$ to reconstruct the data with two combinations of $t+1$ shares, ensuring consistency. It then uses $\phi_2$ to validate the reconstructed data against the description. If the validation fails, the contract refunds all payments for the dataset.  
    \item \textbf{Bad shares from individual \sysname nodes:}  
    If certain shares $\{ds_{i,j}\}$ from DEXO nodes are invalid and cannot reconstruct the data, the consumer invokes the $challenge$ function. The contract checks share integrity via Merkle Tree proofs and applies $\phi_1$ to combine each bad share with $t-1$ valid shares. If the reconstruction fails, the contract flags the share as invalid and refunds payments to the corresponding node.  
\end{enumerate}  
}

\begin{algorithm}[!ht]
\raggedright
\small

\caption{\sysname Contract $\mathcal{C}_{DEXO}$ Pseudocode}
\label{alg:contract}


\KwData{dataSources, seller, buyer, price, $desc$, $\Delta$, $com$, $key$, $keyRevealed$}
\small
\setstretch{1.1}

\SetKwFunction{FMain}{Main}
\SetKwFunction{FConstructor}{constructor}
\SetKwFunction{FInitialize}{initialize}
\SetKwFunction{FBuyerClaim}{query}
\SetKwFunction{FAccept}{accept}
\SetKwFunction{FRevealKey}{revealKey}
\SetKwFunction{FChallenge}{challenge}
\SetKwFunction{FRespond}{respond}
\SetKwFunction{FRefund}{refund}
\SetKwFunction{FNoComplain}{noComplain}
\SetKwFunction{Fphione}{$\phi_1$}
\SetKwFunction{Fphitwo}{$\phi_2$}

\SetKwProg{Fn}{Function}{:}{}
\SetKwProg{InternalFn}{Internal Function}{:}{}

\vspace{2pt}
{\color{gray}\tcc{Smart Contract Constructor}}

\Fn{\FConstructor{DEXO\_Node\_ID, $price$, $desc$, $dataSources[]$, $sellerNodes[]$}}{
    Set $seller\leftarrow DEXO\_Node\_ID$\\
    Set $price, desc, dataSources[], sellerNodes[]$
}

\vspace{2pt}
{\color{gray}\tcc{Initialize data for sale}}

\Fn{\FInitialize{ $\_\Delta$, \_comm}}{
    Require $sender$ in $sellerNodes[]$ {\color{gray} // $sender$: function caller}\\
    Set $\Delta[sender]\leftarrow\_\Delta$ and $commitment[sender]\leftarrow\_comm$\\
}
\vspace{2pt}
{\color{gray}\tcc{Buyer declares itself. 
}}

\Fn{\FBuyerClaim{}}{
    Set $buyer.account \rev{\leftarrow msg.sender}$ \\
    \rev{Set $buyer.desc = \{\_ip,\_port\}$} {\color{gray}//specifying buyer's machine identifier}
}

\vspace{2pt}
{\color{gray}\tcc{
Buyer accepts $z_i$ and transfers payment}}
\Fn{\FAccept{buyer, sellerNode, payment}}{
    Require $payment == price$\\
    $buyer$ deposit for $sellerNode$
}

\vspace{2pt}
{\color{gray}\tcc{Seller reveals the key}}
\Fn{\FRevealKey{key}}{
    Require $sender$ in $sellerNodes[]$\\
    Require $\mathcal{F}_{com}.\mathbf{open}(tid,commitment[\textit{sender}])==\textit{key}$)\\
    Set $keyRevealed[sender] \leftarrow \textit{key}$\\
}

\vspace{2pt}
\rev{
{\color{gray}\tcc{Optional dispute handling}}
\Fn{\FChallenge{casetype, shares1[], shares2[], badShares[], nodeID[], index[], proofs[]}}{
Require $sender$ == $buyer$\\
{\color{gray}// Verify all shares in MTHash Tree}\\
For $j \gets 0$ to $len(nodeID[])$ do:\\
\Indp
$k\leftarrow keyRevealed[\textit{sellerNode[j]}]$ \\
For each $share$ in $shares1[] \cup shares2[] \cup badShares[]$: \\
\Indp
Require $\Delta[nodeID[j]]==\mathbf{MTHash}(\mathbf{Encrypt}_{k}(share), proofs[i], index[i])$;\\
\Indm
\Indm
If $casetype=1$: {\color{gray}// Case 1: Validate Description}\\
\Indp
$d_1 \gets \phi_1(t, n, shares1[])$;\\
$d_2 \gets \phi_1(t, n, shares2[])$;\\
Require $d_1 == d_2$; {\color{gray}// Ensure consistency}\\
Require $\phi_2(d_1, desc)$; {\color{gray}// Validate description}\\
If failed: Refund payments;\\
\Indm
If $casetype=2$: {\color{gray}// Case 2: Verify Bad Shares}\\
\Indp
For $j \gets 0$ to $len(nodeID[])$ do:\\
\Indp
For each $badShare$ in $badShares[]$: \\
\Indp
$d' \gets \phi_1(t, n, \{badShare\} \cup shares1[:-1])$;\\
Require $d' == d_1$;\\
If failed: Refund and mark invalid for badShare;\\
\Indm
\Indm
}
}
\vspace{2pt}
{\color{gray}\tcc{Buyer confirms no complaints}}
\Fn{\FNoComplain{}}{
    Require $sender$ == $buyer$\\
    Transfer funds to sellerNodes and data sources\\
}

\vspace{2pt}
\rev{
\InternalFn{\Fphione{$t, n, S[]$}}{
\Indp
$d \gets \mathcal{F}_{ss}.\mathbf{reconstruct}(t, n, S[])$;{\color{gray} // to instantiate on-chain}\\
Return $d$;\\
\Indm
}
}

\vspace{2pt}
\rev{
\InternalFn{\Fphitwo{$d, desc$}}{
\Indp
Return True if $d$ matches $desc$; False otherwise\\
\Indm
}
}

\end{algorithm}

\subsection{Optimization for On-chain Efficiency}
\label{subsec:optimization}

The design so far requires the consumer server to establish at least $t$ fair exchange sessions with the \sysname network with each session delivering one decryption key. To minimize the on-chain contract execution cost, we introduce two optimizations to allow the consumer to significantly reduce the number of contract calls while retrieving $t$ data shares required for reconstructing the plaintext data. 

\vspace{2pt}
\noindent
\textbf{\rev{Merged} Query and Payment.~} Instead of having the consumer query each node (step \circled{1}) and make payments according to each exchange session (step \circled{3}) separately, they can batch-process the exchange sessions with all nodes in one contract call. That is, step \circled{1} is now a query on the shares from all $N$ nodes; step \circled{3} is now a payment for all queried shares. 

\vspace{2pt}
\noindent
\textbf{Shared Key for $(t-F)$ Nodes.~} This method aims to minimize the number of fair exchange sessions without hampering the final delivery of requested data shares. Instead of having every node $i$ generate a new share-encryption key $k_i$ for its exchange with the consumer, the P-DApp server may pre-select $(t-F)$ nodes as the ``priority group'' with a certain node serving as the group leader (denoted $\mathcal{N}_p$) who will coordinate the generation of a common secret key $k_p$ for the entire group (group key generation has been well studied and is thus orthogonal to our work).
In this way, the consumer can invoke one exchange session with $\mathcal{N}_p$ to obtain the corresponding $k_p$ for the data shares held by all $(t-F)$ nodes in this group (Step \circled{2} still need to be performed on each node). The consumer only needs to exchange with other $F$ nodes to obtain the remaining $F$ shares. In this way, the total number of fair exchange sessions has been reduced from $t$ to $F+1$. If the system is configured $N\geq3F+1$ and $t=\frac{2}{3}N$, this marks at least a 50\% reduction in the sessions. We will show in Section \ref{subsec:informal-analysis} that \sysname is still secure when this optimization is applied.

\begin{figure}
    \centering
    \includegraphics[width=\linewidth]{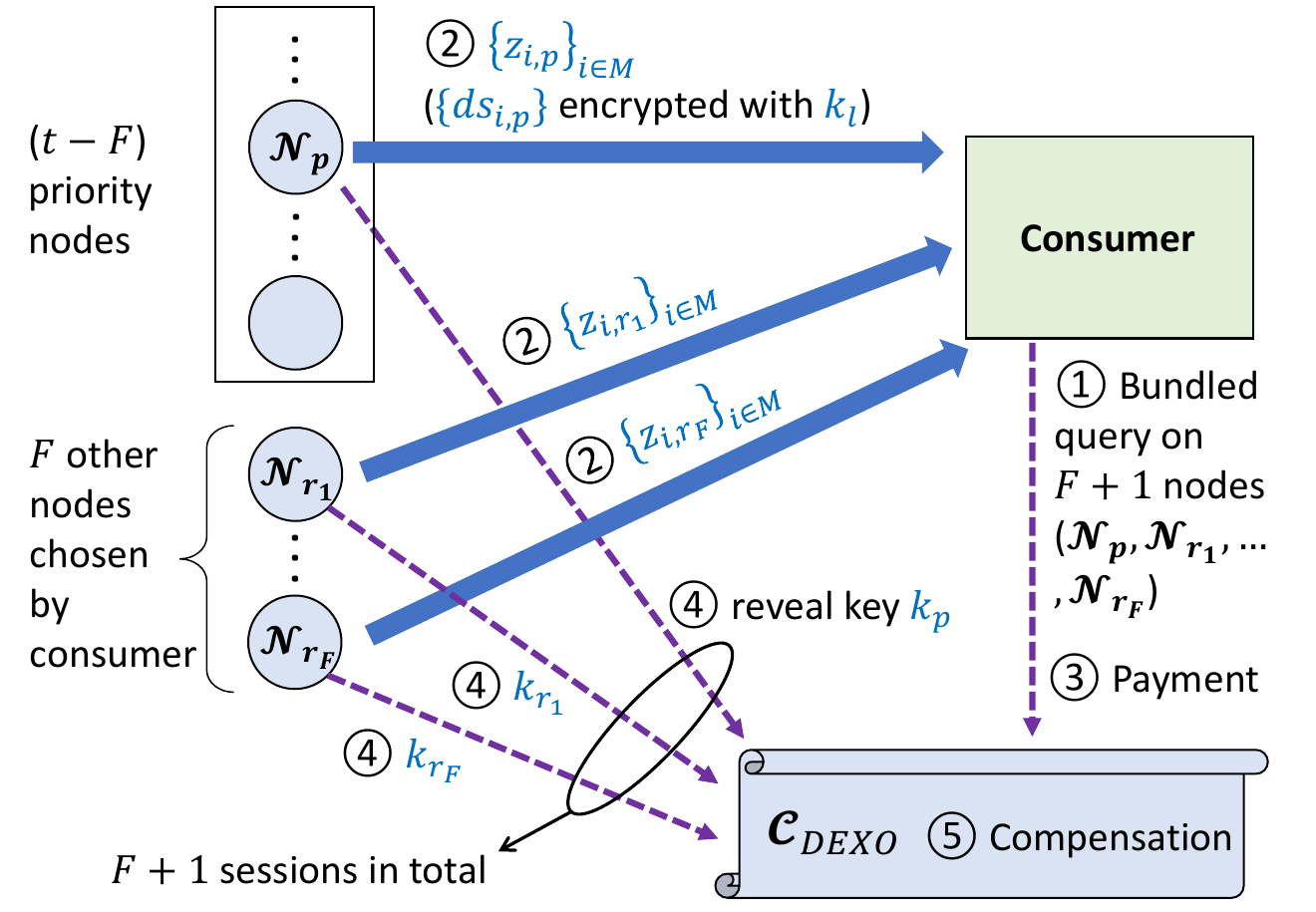}
    \caption{Fair exchange process with \rev{merged query} and shared key  (as described in \S\ref{subsec:dexo-contract} and \S\ref{subsec:optimization}). A total of $F+1$ parallel exchange protocol sessions are required.}
    \label{fig:fair-exchange}
\end{figure}

The overall procedure after the two optimizations are applied is shown in Fig. \ref{fig:fair-exchange}.

\section{Analyses}
\label{sec:analyses}

\subsection{Security Analysis}
\label{subsec:informal-analysis}

In this section, we show that \sysname fulfills the proposed security goals under the defined threat model (Section \ref{subsec:threat-model}).

\begin{theorem}[Data Source Verifiability]
Each data source, i.e., a P-DApp user device, always performs data pre-processing, secret sharing, and signing as specified in the trusted application $\mathcal{F}_{TA}$ correctly. The execution integrity can be verified later by a \sysname node.
\end{theorem}
\begin{IEEEproof}
Establishing the trusted application $\mathcal{F}_{TA}$ as a TEE instance on a P-DApp user device ensures its execution integrity. The correctness of this process reduces to the integrity of TEE-based attested execution $\mathcal{G}_{att}$ as described in Section \ref{subsec:buildingblock-tee}.
Moreover, each signature within $\sigma_{i,j}$ ($j\in[N]$) generated by user $i$'s $\mathcal{F}_{TA}$ instance proves the integrity of both the TEE runtime environment and the generated data share $ds_{i,j}$ which is verifiable by \sysname Nodes with the help of the TEE attestation service \rev{that verifies} the validity of the user device's TEE public key. 
\end{IEEEproof}

\rev{This mechanism ensures that even if an adversary gains control over a P-DApp server or other parts of the system, they cannot tamper with the data processing or generate fraudulent data shares without detection. The TEE attestation process guarantees that only data shares produced by genuine, unaltered TEE instances are accepted by the DEXO nodes, effectively mitigating spoofing or data manipulation attacks.}

\begin{theorem}[End-to-end Confidentiality]
\label{theorem:confidentiality}
The formatted data generated by a P-DApp user device $i$, $d_i$, is only delivered to the paid consumer while being hidden from other parties, including individual \sysname nodes and other consumers.
\end{theorem}
\begin{IEEEproof}
This property derives from the confidentiality of TEE-based attested execution $\mathcal{G}_{att}$ and the security of secrete sharing functionality $\mathcal{F}_{ss}$. More specifically, $\mathcal{G}_{att}$ generates $d_i$ within the TEE and the $(t,N)$-secret sharing outputs only the data shares $\{ds_{i,j}\}$. Since we assume $F<t$, the compromised \sysname nodes cannot obtain enough data shares to reconstruct $d_i$. The same applies to malicious consumers who do not pay to receive decryption keys for at least $t$ data shares of $d_i$. \rev{Therefore, \sysname prevents malicious nodes or non-paying consumers from reconstructing the original data the confidential data in the sale.}
\end{IEEEproof}


\begin{theorem}[Fault Tolerance of Data Delivery]
\label{theorem:fault-tolerance}
The consumer is guaranteed to receive the requested formatted data $d_i$ despite the presence of up to $F$ compromised \sysname nodes.
\end{theorem}
\begin{IEEEproof}
It is sufficient to show that the consumer can always receive the data shares to reconstruct the data. Since we assume $F<\frac{1}{2}N$, thus $N-F>F$. Therefore, there always exists a $t$ so that $F< t \leq N-F$. This guarantees that the consumer can receive at least $N-F$ valid data shares from non-faulty nodes which can be used to reconstruct $d_i$ (since $t \leq N-F$).
\end{IEEEproof}

We remark that Theorem \ref{theorem:confidentiality} and Theorem \ref{theorem:fault-tolerance} imply the decentralization property since the system is resilient to single-point failure among \sysname nodes. This prevents any single or minority nodes from monopolizing or leaking the data.

\begin{theorem}[Fair Exchange with Collusion Resistance]
The P-DApp users receive compensation only if the consumer obtains the correct data. Simultaneously, the consumer receives the data only if the P-DApp users receive the requested compensation. This process is secure even if either party colludes with compromised \sysname nodes.
\end{theorem}
\begin{IEEEproof}
We first show the security against source-node collusion. Consider the worst-case scenario when $F$ \sysname nodes are compromised and willing to collude with the P-DApp to scam a consumer without fulfilling the data provision. The $F$ nodes can send crafted data shares to the consumer. Since the consumer receives at least $t$ shares and $t>F$, it can always detect the inconsistencies of the received shares and invoke the dispute protocol (Section \ref{subsec:dexo-contract}) to abort the entire exchange and get the payment back. When the two optimizations (see \S\ref{subsec:optimization}) are applied, the consumer still gets at least $F+1$ shares from each user; there is at least one correct share for the consumer to uncover the inconsistencies of the received shares. This provides sufficient evidence for the consumer to start a dispute process and claim the payment back (see \S\ref{subsec:dexo-contract}).

Next, we show the security against consumer-node collusion. Consider the worst-case scenario when the $F$ compromised \sysname nodes collude with the consumer to trick a P-DApp into providing the correct data without a full payment. The $F$ nodes can send all their shares to the consumer, which however is not sufficient for the latter to reconstruct the original data since $F<t$. When the two optimizations (Section \ref{subsec:optimization}) are applied, we consider the worst-case scenario when only one of the nodes with shared keys is compromised---this will make the data shares from all the $t-F$ nodes available to the consumer. However, counting in the remaining $F-1$ compromised nodes, the consumer can still obtain at most $t-F+F-1=t-1$ shares, below the required $t$ shares. Knowing the $t-1$ shares is no different from knowing one share since they both reveal no useful information about the original data.

\rev{Lastly, in case a consumer starts a dispute process, the same collusion resistance is achieved. This is shown in Lemma \ref{lemma:integrity-dispute-handling} below.}
\end{IEEEproof}

\rev{
\begin{lemma}[Integrity of Dispute Handling]  
\label{lemma:integrity-dispute-handling}
    The dispute handling process of \sysname is secure even if either party colludes with compromised \sysname nodes.  
\end{lemma}  
\begin{IEEEproof}  
    The security of the dispute-handling process relies on the integrity of the secret-sharing scheme and the predicate functions $\phi_1$, $\phi_2$. Even if a subset of DEXO nodes (\(F < t\)) colludes with one party (e.g., a consumer or a P-DApp), they cannot tamper with or fabricate valid data shares without detection.  
    During a dispute, the contract first verifies the authenticity of all shares to ensure they originate from valid Merkle Tree roots. This step prevents a malicious consumer from injecting fake shares into the contract. After verifying the shares, the contract uses $\phi_1$ to reconstruct the original data multiple times with different combinations of shares, ensuring consistency. It then applies $\phi_2$ to validate that the reconstructed data matches the description specified in the contract. Any mismatch triggers a dispute and refunds payments.  
    Furthermore, the contract checks each suspected bad share by combining it with $t-1$ valid shares and reconstructing the data using $\phi_1$. If the reconstructed data does not match the verified original data, the share is flagged as invalid, and payments are refunded.  
    The TEE attestation mechanism ensures that all data shares originate from trusted, unaltered TEE instances. Even if a consumer colludes with compromised nodes, they cannot generate sufficient valid shares to reconstruct the data without detection, as \(t\) valid shares are required. Similarly, a colluding P-DApp cannot cheat a consumer by providing incorrect data, as integrity checks during reconstruction will fail.  
    Thus, \sysname's dispute handling guarantees fairness and integrity in resolving conflicts, even in the presence of malicious behaviors or collusion.  
\end{IEEEproof}
}

\subsection{On-chain Complexity Analysis}

Here we analyze how the number of P-DApp users $M$ and the number of \sysname nodes $N$  may affect the on-chain complexity.
\rev{In our analysis, the optimal scenario assumes that all parties, including the buyer and DEXO nodes, act honestly according to the protocol. In this scenario, the transaction proceeds directly to completion without any disputes.}
\rev{
In this optimal scenario, the P-DApp server constructs the contract and initializes the data exchange process. Each of the $N$ DEXO nodes performs an Initialize operation to register metadata, such as Merkle root hashes, with the contract. The consumer submits a single query operation to request all necessary data shares, after which the $N$ nodes verify the query and transmit encrypted data shares off-chain to the consumer. The consumer then makes $t$ payments, corresponding to the $t$ threshold shares required to reconstruct the original data. The $N$ nodes verify payments, and $t$ nodes detect and confirm receipt of payments. These $t$ nodes proceed to execute the revealKey operation, releasing encryption keys that the consumer retrieves through $t$ get key operations. Altogether, this process involves a total of $3N + 3t + 2$ contract calls, which are independent of the number of users ($M$) since the P-DApp server aggregates and packages data shares before interacting with the contract.
}

\rev{Intuitively, the irrelevance between $M$ and the total number of contract calls is due to the fact that individual users do not interact with the contract by themselves. Instead, each \sysname node $j$ collects the $j$th data share from all users and registers their cryptographic digest via one $initialize$ function call. The same applies to the disclosure of decryption keys, where node $j$ makes one $revealKey$ function call to disclose $k_j$ for decrypting all users' shares. When a fair exchange process finishes, the consumer invokes the contract via one $noComplain$ call to dispute payments to multiple users at once. These designs avoid the sheer amount of calls from data sources.} \revtwo{Our system design reduces direct blockchain interactions. By delegating data storage and management to DEXO nodes, the design offloads resource-intensive processes, significantly reducing on-chain overhead.}

\rev{However}, the on-chain complexity also needs to take into account the blockchain execution cost (e.g., denominated in gas fee in Ethereum). \rev{In one fair exchange, the gas cost of calling} $noComplain$ is dependent on the number of users $M$ \rev{since the smart contract allocates payments to the users through $M$ internal transactions. Even though an internal transaction requires minimal gas cost compared to a normal contract, the total gas consumption is still linear in $M$.}
We will demonstrate the gas cost of contract execution \rev{$noComplain$'s $M$-dependent cost} in Section \ref{Compare Gas Cost of DEXO with Chainlink}. \revtwo{Specifically, the gas cost of this function scales linearly with \( M \), requiring approximately 5735 units of gas for each additional user. This translates to an incremental cost of around \$0.22 per user under current Ethereum gas prices (at May 2024 Ethereum price).} \rev{If a dispute happens, the gas cost of calling the $Challenge$ function is also dependent on $M$ since the proof size submitted by a consumer may be proportional to the size of data shares received from one \sysname node.} 


\rev{Lastly, when the blockchain platform experiences congestion due to a high volume of transactions from external activities, \sysname consumers may still encounter increased transaction latency and elevated gas fees due to network competition, similar to all Web3 users in general. Addressing these fundamental limitations would require scalability solutions for the blockchain platform itself, such as integrating sidechains, payment channels or other Layer-2 mechanisms, or switching to private blockchains exclusive to our system. These considerations are of practical importance and we leave them to future study.}

\section{Implementation}
\label{sec:implementation}
We provide a proof-of-concept implementation of \sysname's system components, including the TEE-capable P-DApp user, \sysname node, data consumer, and the \sysname contract.\footnote{Our code is available at \url{https://github.com/yli568/DEXO}.}


We used Raspberry Pi 3 (RPi3) to simulate an IoT/mobile data provider (\textit{i.e.}, a P-DApp user device) and OP-TEE \cite{OP-TEE-on-Github} to implement the TEE-based attested execution functionalities based on RPi3's native ARM TrustZone support \cite{ARMTrustZone} and Open Portable TEE (OP-TEE) \cite{OP-TEE-on-Github}. OP-TEE is an open-source implementation of the TEE concept primarily targeting ARM-based devices and is designed to provide a secure and isolated environment for running sensitive code and processing confidential data. The OP-TEE project can be compiled into the Linux system which can be run on the RPi3 board.
We utilized the Repo Manifest \cite{Repo-manifest-for-OP-TEE} to compile and configure the various components involved.
For the TrustZone TEE attestation, we utilized the attestation function of OP-TEE released in April 2022 ~\cite{optee-attestation-released}. 
For the secret-sharing-based data generation procedure, we ported an off-the-shelf implementation \cite{fletcher2023csss} of Shamir’s
Secret Sharing~\cite{shamir1979share} into the TEE program $\mathcal{F}_{TA}$. The native RSA signature function is used to generate a signed attestation report at the end of $\mathcal{F}_{TA}$.
\textbf{\sysname Contract.~}
We implemented  a proof-of-concept $\mathcal{C}_{\sysname}$, which is shown in Algorithm \ref{alg:contract}, with Solidity for the Ethereum blockchain containing about 130 lines of code. It realizes the important functions except the optional dispute handling routine described in Section \ref{subsec:dexo-contract}. The contract was deployed to the Ethereum Sepolia testnet for evaluating the gas and time cost of running the data exchanges.
\section{Evaluation}
\label{sec:evaluation}

\label{subsec:eval-criteria}



We conducted experiments under varying conditions of \sysname to evaluate the following performance metrics:
(i) Time and gas fee costs associated with using \sysname to obtain data as a data consumer, and its comparison to existing approaches.
(ii) Time cost for a data provider to generate data in the TEE environment.
The above metrics are evaluated for scalability under different number of data providers and \sysname nodes.

\subsection{Time Cost of Transaction}

\begin{figure}
    \centering
    \includegraphics[width=\linewidth]{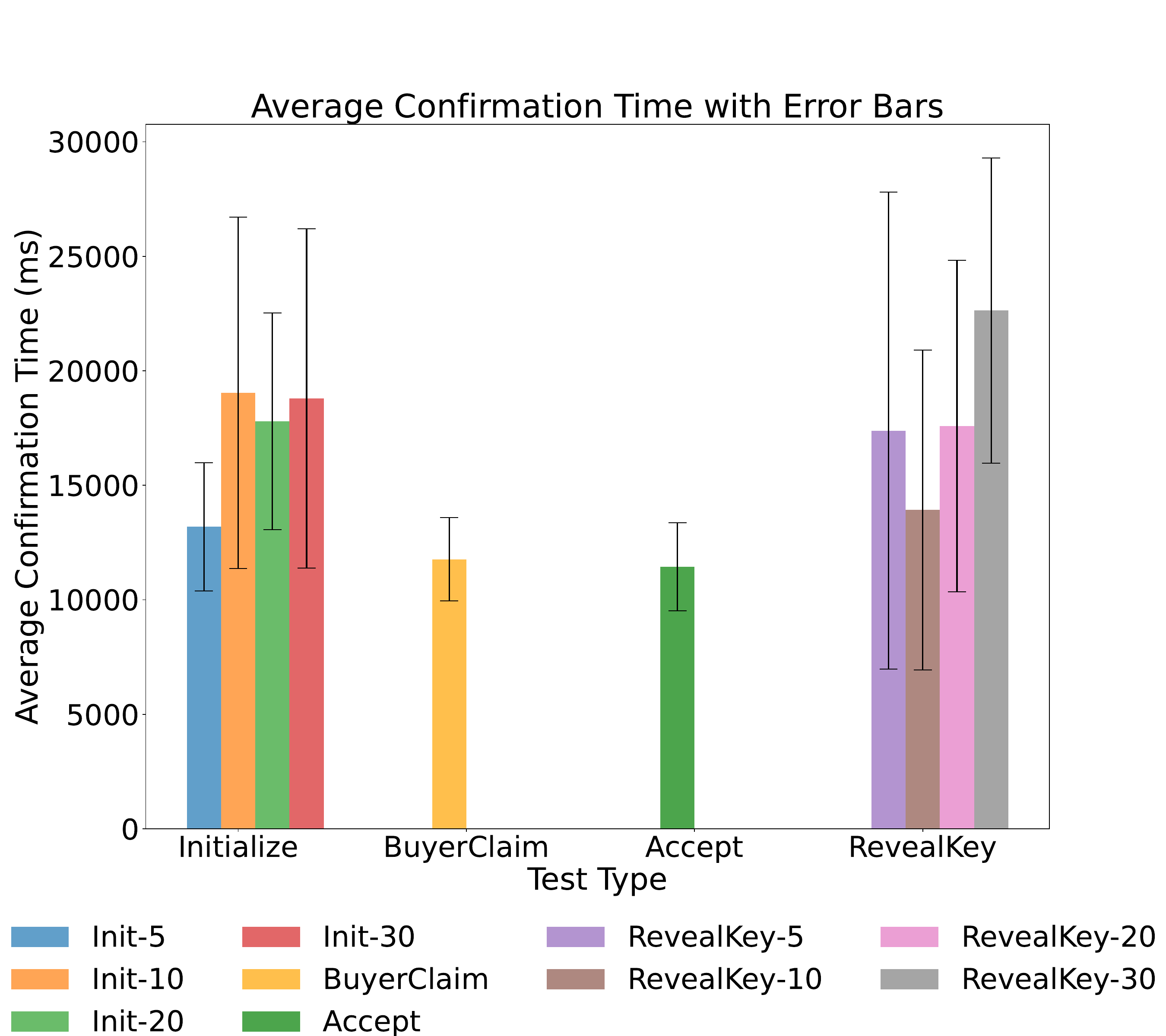}
    \caption{Average Confirmation Times. Note: The suffix number represents the number of nodes used in the test. Those nodes call this function to the contract on the testnet at the same time}
    \label{fig:average_confirmation_times_with_error_bars_adjusted}
\end{figure}

Fig.~\ref{fig:average_confirmation_times_with_error_bars_adjusted} shows the time costs of invoking $\mathcal{C}_{\sysname}$ over the Sepolia testnet. It measures the average confirmation times (in milliseconds) for different smart contract functions: $initialize$, $buyerClaim$ (the equivalent of the \sysname contract's $query$ function), $accept$, and $revealKey$. Each test type is represented with a different color bar, and error bars indicate the standard deviation in the measurements.

For a buyer seeking to retrieve desired data after discovering it, the performance of the $buyerClaim$, $accept$, and $revealKey$ functions are particularly relevant. $buyerClaim$ and $accept$ have average confirmation times of approximately 12.5s. And $revealKey$ has average confirmation times of approximately 18s with a \sysname network with 20 nodes. This suggests that, under our test condition, a buyer can expect to complete these transactions and retrieve the data within 1 minute.

The extended confirmation times for the RevealKey function can be attributed to the complex validation procedures within the function. Specifically, the function includes checks to determine the validity of the provided key, which requires additional computational effort. This increased processing time on the testnet leads to longer intervals before a block can be confirmed. It is important to note that the computation results of a call are obtained before the block confirmation, not after.

\subsection{Gas Cost of DEXO}

Execution of $\mathcal{C}_{DEXO}$ incur costs in terms of computational resources, which are quantified as gas costs in the Ethereum network.
We evaluate the gas costs (on-chain execution fees) associated with various $\mathcal{C}_{DEXO}$ operations as shown in Table ~\ref{tab:gas_costs}.
Firstly, deploying the DEXO smart contract is a \textit{one-time} operation that sets up the contract on the Ethereum network. This computationally intensive operation results in gas consumption of approximately 2,325,998 units. In May 2024, with a gas price of 10.96 gwei and the value of Ether at \$3,510 (USD), the approximate cost is about \$89.48. However, it is essential to note that due to the volatile nature of Ether price and gas price on the Ethereum network, this cost can significantly fluctuate in real-world conditions.

The ``initialize" function has a gas fee of 74,248 units (\$2.85). The ``noComplain'' function is used for distributing the revenue. The base gas fee for invoking this function is 37,194 units (\$1.43). In addition to this base fee, an extra 5,735 gas units (\$0.22) are required for each data source included in the distribution.
The gas fees for the ``Accept'', ``revealKey'', and ``check Key'' functions in Solidity are fixed at 74,843, 84,334, and 3,457 units (i.e., \$2.87, \$3.24, and \$0.13), respectively, and do not vary with the number of elements in the dataSources array.

\begin{table}[ht]
\centering
\caption{Gas Costs of Invoking $\mathcal{C}_{DEXO}$ Functions}
\label{tab:gas_costs}
\begin{tabular}{@{}lS[table-format=7.0]S[table-format=5.2]@{}}
\toprule
\textbf{Function}    & \textbf{gas fee (Units) \& Cost (USD)}\\
\midrule
Deployment          & {2,325,998 (\$89.48)}\\
Initialize           & {74,248(\$2.85)}\\
noComplain           & {37,194 (\$2.05) + 5,735 (\$0.22) $\times$ \#DSs}\\
Accept               & {74,843 (\$2.87)}\\
revealKey            & {84,334 (\$3.24)}\\
check Key             & {3,457 (\$0.13)}\\
\bottomrule
\end{tabular}

\end{table}



\subsection{Comparing Gas Cost of DEXO with Chainlink}
\label{Compare Gas Cost of DEXO with Chainlink}

We compare the on-chain gas cost with the popular data oracle solution Chainlink \cite{breidenbach2021chainlink}, which delivers data through contract API calls. According to \cite{khan2022investigation}, the gas fee for a ``Price Feed'' transaction is 216,844 units (\$8.34), while an API Call incurs a gas fee of 1,470,295 units (\$56.56). 
A single call through Price Feeds or an API Call on Chainlink or its Oracles typically retrieves a singular data point, such as the current weather condition at a specific time, the real-time price of a cryptocurrency, or the current market value of a specific stock. In comparison, \sysname delivers the ciphertext data off-chain and only uses blockchain for fair exchange transactions, significantly reducing the on-chain gas cost.

\begin{figure}
    \centering
    \subfigure[]{
        \hspace{-20pt}
        \includegraphics[width=0.275\textwidth]{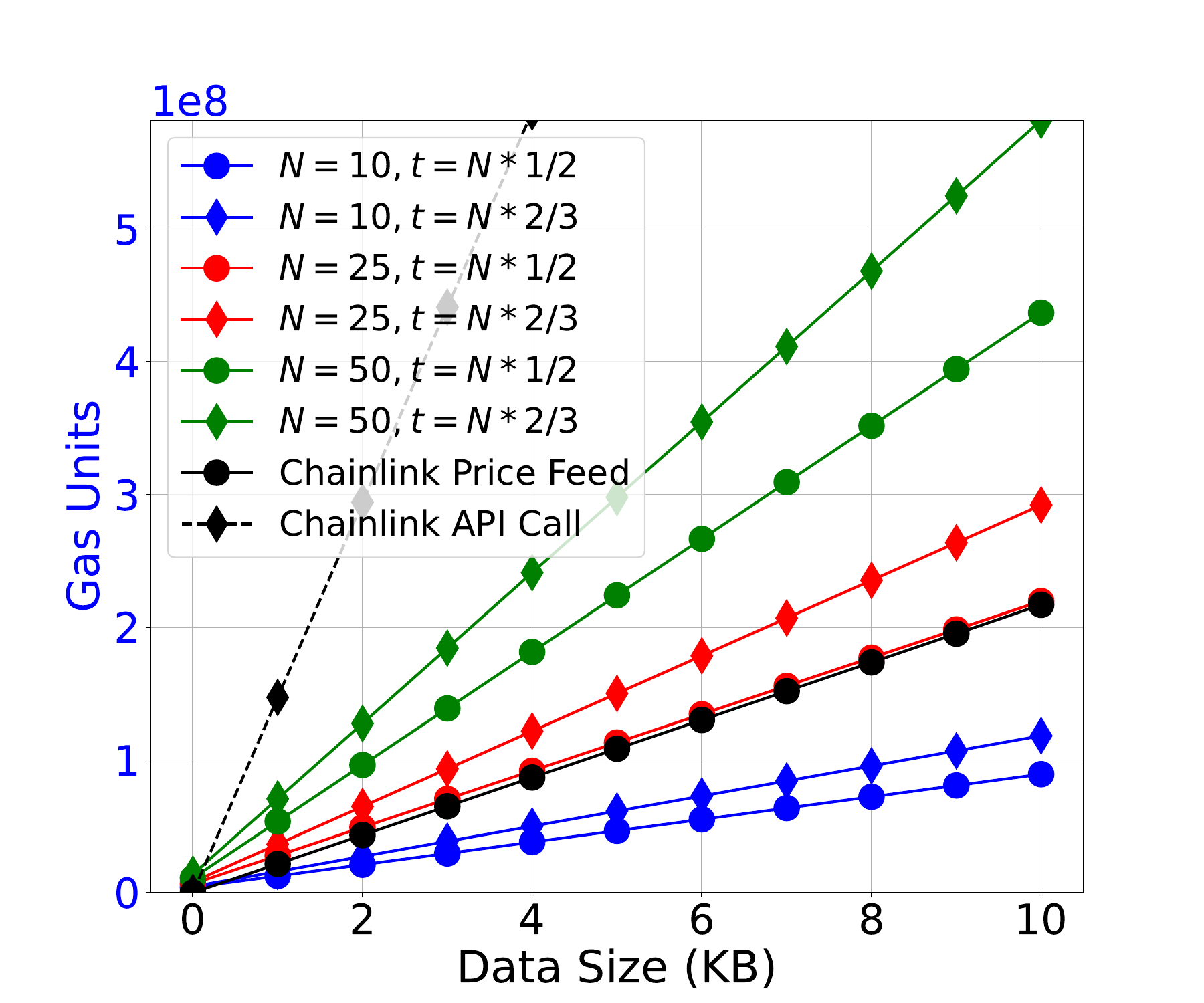}
        \label{fig:Compare Gas Cost of DEXO with Chainlink - 10B}
        \hspace{-15pt}
    }
    \subfigure[]{
        \hspace{-15pt}
        \includegraphics[width=0.275\textwidth]{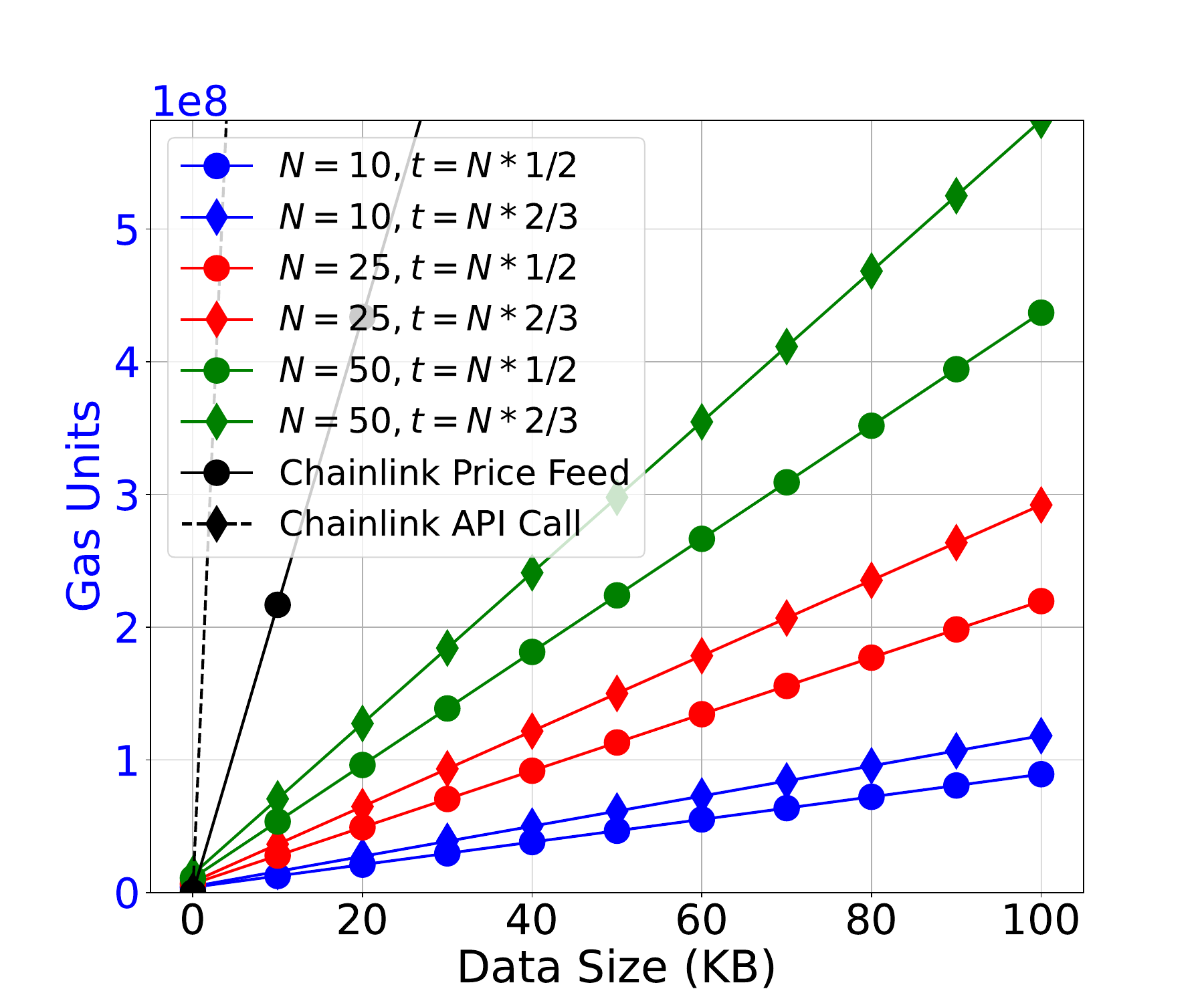}
        \label{fig:Compare Gas Cost of DEXO with Chainlink - 100B}
        \hspace{-20pt}
    }
    \caption{Compare Gas Cost of DEXO with Chainlink. (a) Each user provides each share of a one-time 10B data to each Node. (b) Each user provides each share of a one-time 100B data to each Node.}
    \label{fig:gas-cost}
\end{figure}

Based on the above benchmarking result, we provide an extrapolation analysis of on-chain gas costs when we scale up the number of data providers and compare them to Chainlink 
The results are shown in Fig. \ref{fig:Compare Gas Cost of DEXO with Chainlink - 10B} and Fig. \ref{fig:Compare Gas Cost of DEXO with Chainlink - 100B}. 
As the number of \sysname nodes increases, the gas fee increases linearly because the buyer must transact with more nodes to acquire a sufficient number of shares for aggregating usable data. When $t=\frac{2}{3}n$, the gas fee incurred is higher than that when $t=\frac{1}{2}n$ because a smaller threshold signifies the need for fewer shares to aggregate usable data, thereby implying fewer transactions.

Specifically, Fig. \ref{fig:Compare Gas Cost of DEXO with Chainlink - 10B} represents the scenario where each P-DApp user contributes 10 bytes of data per instance. An increase in Data Size implies an increase in the number of users. For the two methods of Chainlink, an increase in Data Size signifies a rise in the number of requests made to Chainlink. The result shows that if each P-DApp user contributes a small amount of data per instance, the gas fee for Chainlink Price Feed is roughly equivalent to the gas fee for \sysname when $n=25$ and $t=\frac{1}{2}n$. Moreover, all the scenarios for \sysname listed outperform the Chainlink API Call regarding gas fees.
Fig. \ref{fig:Compare Gas Cost of DEXO with Chainlink - 100B} depicts the scenario when each P-DApp user provides 100 bytes of data per instance. This figure illustrates that if each P-DApp user contributes slightly larger amounts of data per instance, all scenarios for \sysname outperform the Chainlink Price Feed. This further highlights \sysname's significant advantage in transmitting larger volumes of data from an increasing number of data providers.





\subsection{TEE Overhead}

We evaluated the computation overhead attributed by the ARM TrustZone TEE on the P-DApp user end. We tested the TEE program $\mathcal{F}_{TA}$ to generate data, shares, and sign a single share with the runtime environment. 
Based on the testing conducted with OP-TEE on RPi3, it has been observed that generating shares in TEE and signing them with the TA execution environment costs variant based on the size of the original data and the number of shares that \sysname needs, which is the number of nodes in the \sysname network.

\begin{figure}
    \centering
    \includegraphics[width=.7\linewidth]{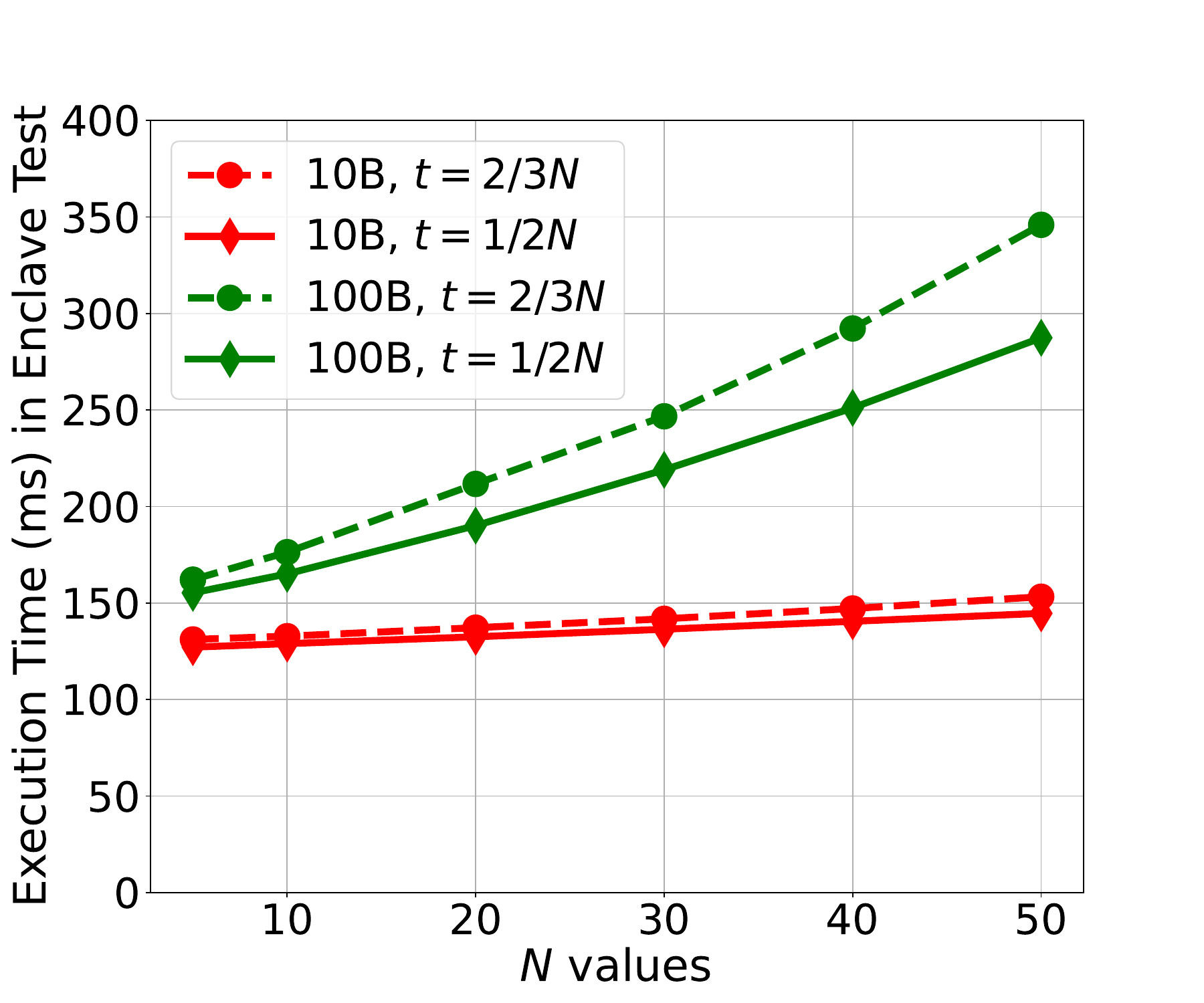}
    \caption[TSecret Sharing Results]{Time cost of data provider's TEE operation ($\mathcal{F}_{TA}$). Here `10B' refers to the size of the original data involved in the secret sharing process, which is 10 bytes. Similarly, `100B' indicates that the original data size is 100 bytes.}
    \label{fig:Enclave_Test_plot_modified}
\end{figure}

Fig.~\ref{fig:Enclave_Test_plot_modified} shows the execution time (ms) in an enclave test for varying node counts (N) in a \sysname network, with N values ranging from 10 to 50. The y-axis ranges from 0 to 400 ms.
For 10B origin data (red lines), we observe that (i) the execution time increases relatively slowly across all N values and (ii) the difference between thresholds $t=N/2$ (solid) and $t=2N/3$ (dashed) is not significant.
For 100B origin data (green lines), we observe that (i) the execution time increases with the number of nodes faster and (ii) the threshold $t=N/2$ (solid) is consistently lower than $t=2N/3$ (dashed).

Overall, execution time increases with larger data sizes and higher threshold values, especially as the number of nodes grows. In practice, we can assume a fixed number of $N$, similar to existing DON solutions (e.g., Chainlink), and a maximum size of data entry. In future work, we will explore more efficient TEE-based secret-sharing implementations to reduce the data provider's operational cost further.

\section{Conclusion}
\label{sec:conclusion}

In this work, we introduced a new decentralized data exchange mechanism called \sysname for enabling a secure and scalable marketplace for IoT and mobile data. By augmenting the decentralized oracle network paradigm with innovative hardware-cryptographic co-design that harmonizes trusted hardware, secret sharing, and blockchain smart contract, \sysname for the first time enables secure data exchange between distributed data providers and consumers while fulfilling end-to-end data confidentiality, source verifiability, decentralization, and fairness goals with strong resilience to participant failures and collusions. The experiment results demonstrate \sysname's feasibility in the deployment with Ethereum smart contracts with moderate on-chain gas cost per unit of data consumed while incurring minimal off-chain execution overhead for individual data providers. 

\section*{Acknowledgments}
This work was supported in part by the US National Science Foundation under grant numbers 2247561 and 2238680, and by the Office of Naval Research under grant number N00014-24-1-2730 subawarded through Virginia Tech.



\bibliographystyle{ieeetr}
\bibliography{reference,reference-xiao}

\begin{IEEEbiography}[{\includegraphics[width=1in,height=1.25in,clip,keepaspectratio]{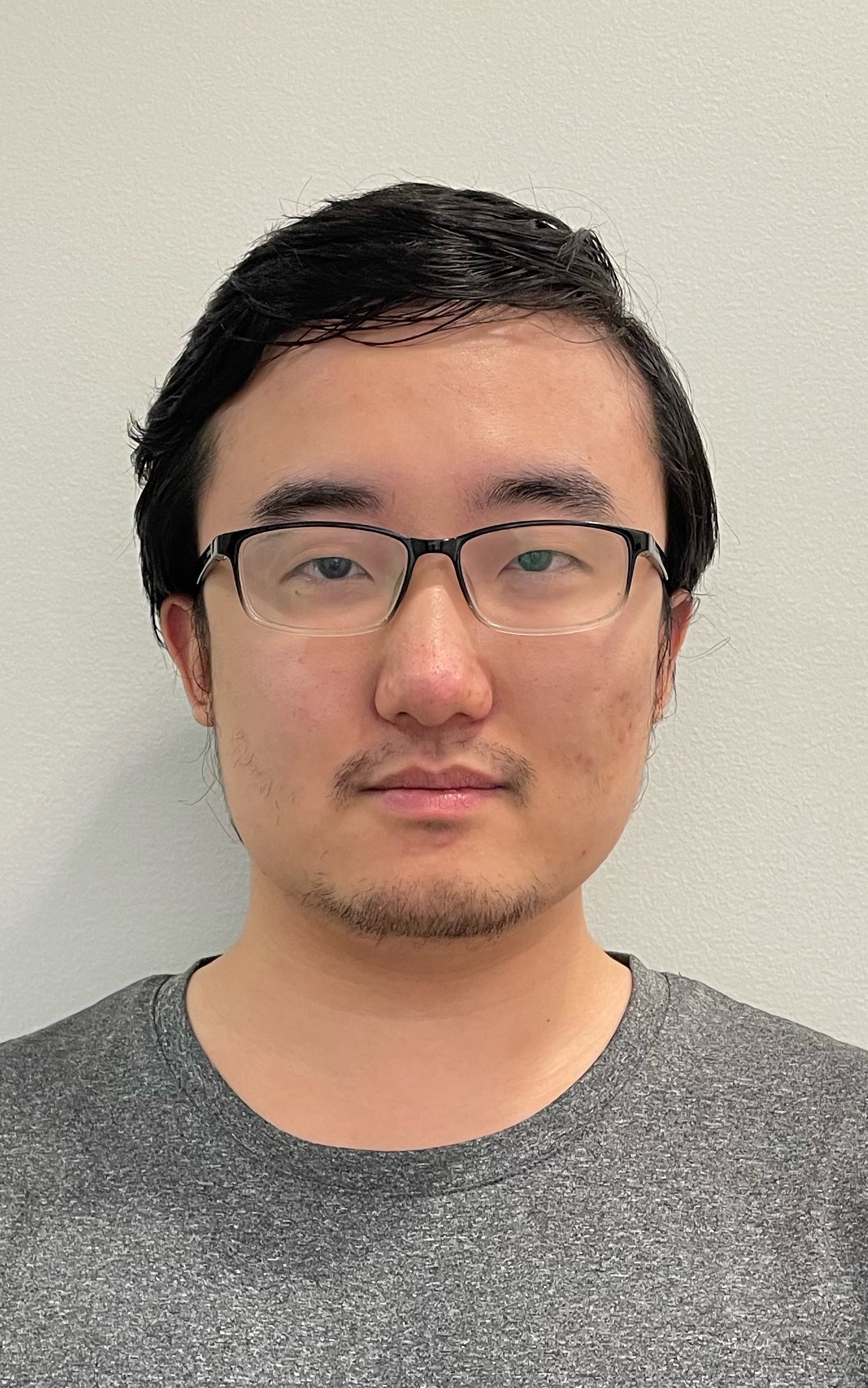}}]{Yue Li} (Student Member, IEEE) received the M.Sc. degree in Computer Science from the Department of Computer Science, University of Kentucky, Lexington, KY, USA, in 2023 and is currently a Computer Science Ph.D. student in the same department. He received his bachelor's degree in Software Engineering from the University of Electronic Science and Technology of China, Chengdu, Sichuan, China. His research interests lie in distributed system security and decentralized systems.
\end{IEEEbiography}

\begin{IEEEbiography}[{\includegraphics[width=1in,height=1.25in,clip,keepaspectratio]{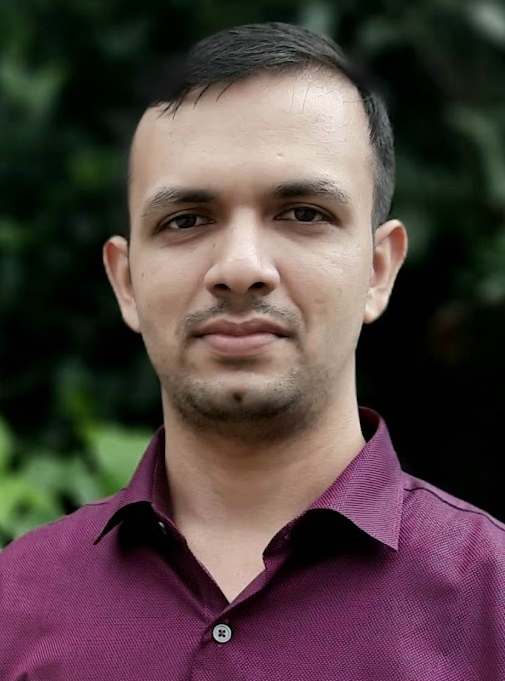}}]{Ifteher Alom} (Student Member, IEEE) is a Computer Science graduate student at the University of Kentucky, Lexington, KY, USA. He received his bachelor's degree from the Department of Computer Science, Shahjalal University of Science and Technology, Sylhet, Bangladesh. His research interests include applied cryptography, identity management, and blockchain applications.
\end{IEEEbiography}

\begin{IEEEbiography}[{\includegraphics[width=1in,height=1.25in,clip,keepaspectratio]{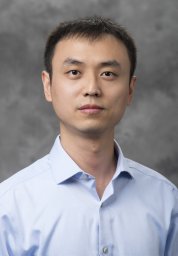}}]{Wenhai Sun} (Senior Member, IEEE) holds
two Ph.D. degrees from the School of
Telecommunications, Xidian University, and from
the Department of Computer Science, Virginia
Tech, respectively.
He is currently an Assistant Professor with
the Department of Computer and Information
Technology, Purdue University, West Lafayette, IN,
USA, and a Faculty Member affiliated with the
Center for Education and Research in Information
Assurance and Security at Purdue. His research interests lie in privacy-enhancing technologies, confidential computing, AI/ML security, decentralized trust and infrastructure, and cyber-physical systems security.
Dr. Sun won the Distinguished Paper Award in ACM ASIACCS 2013. He received the NSF CAREER Award in 2023.
\end{IEEEbiography}

\begin{IEEEbiography}[{\includegraphics[width=1in,height=1.25in,clip,keepaspectratio]{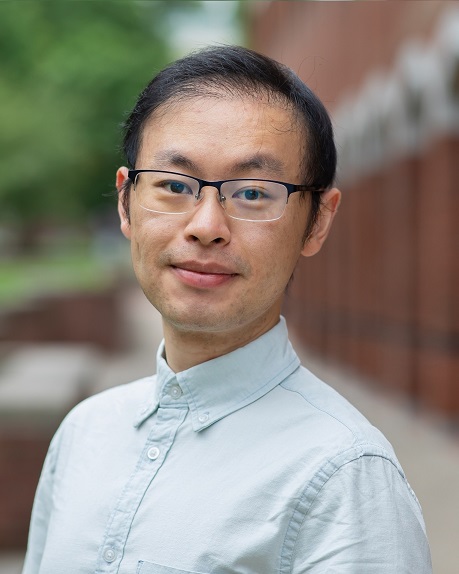}}]{Yang Xiao} (Member, IEEE) received the Ph.D. degree in Computer Engineering from the Department of Electrical and Computer Engineering, Virginia Tech, USA in 2022. He is currently an Assistant Professor with the Department of Computer Science at the University of Kentucky, Lexington, KY, USA. His research interests lie in network security,  distributed system security, blockchain and decentralized systems, and mobile network security. 
\end{IEEEbiography}

 




\vfill

\end{document}